\documentclass[twocolumn]{aastex701}
\shorttitle{}
\shortauthors{}
\usepackage{longtable, booktabs, multirow}
\usepackage{graphicx} 
\usepackage{array}  
\usepackage{amsmath}
\usepackage{booktabs}
\usepackage{enumitem}
\usepackage{rotating}
\usepackage{mathtools, multirow, tabularx, threeparttable}
\usepackage{CJK}
\usepackage[utf8]{inputenc}
\usepackage{etoolbox} 

\begin{document}

\title{A JWST/MIRI Study of Dust in a Sample of Normal Type IIP Core Collapse Supernovae}


\newcommand{\LCO}{\affiliation{Las Cumbres Observatory, 6740 Cortona Drive, Suite 102, Goleta, CA 93117-5575, USA}}
\newcommand{\UCSB}{\affiliation{Department of Physics, University of California, Santa Barbara, CA 93106-9530, USA}}
\newcommand{\KITP}{\affiliation{Kavli Institute for Theoretical Physics, University of California, Santa Barbara, CA 93106-4030, USA}}
\newcommand{\UCD}{\affiliation{Department of Physics and Astronomy, University of California, Davis, 1 Shields Avenue, Davis, CA 95616-5270, USA}}
\newcommand{\WIS}{\affiliation{Department of Particle Physics and Astrophysics, Weizmann Institute of Science, 76100 Rehovot, Israel}}
\newcommand{\OKC}{\affiliation{Oskar Klein Centre, Department of Astronomy, Stockholm University, Albanova University Centre, SE-106 91 Stockholm, Sweden}}
\newcommand{\OAPD}{\affiliation{INAF-Osservatorio Astronomico di Padova, Vicolo dell'Osservatorio 5, I-35122 Padova, Italy}}
\newcommand{\UniPd}{\affiliation{Dipartimento di Fisica e Astronomia ``G. Galilei'', Universit\`a degli studi di Padova Vicolo dell'Osservatorio 3, I-35122 Padova, Italy}}
\newcommand{\CaltechCahill}{\affiliation{Cahill Center for Astronomy and Astrophysics, California Institute of Technology, Mail Code 249-17, Pasadena, CA 91125, USA}}
\newcommand{\Caltech}{\affiliation{Department of Astronomy and Astrophysics, California Institute of Technology, Pasadena, CA 91125, USA}}
\newcommand{\GSFC}{\affiliation{Astrophysics Science Division, NASA Goddard Space Flight Center, Mail Code 661, Greenbelt, MD 20771, USA}}
\newcommand{\UMD}{\affiliation{Joint Space-Science Institute, University of Maryland, College Park, MD 20742, USA}}
\newcommand{\UCB}{\affiliation{Department of Astronomy, University of California, Berkeley, CA 94720-3411, USA}}
\newcommand{\TTU}{\affiliation{Department of Physics, Texas Tech University, Box 41051, Lubbock, TX 79409-1051, USA}}
\newcommand{\STScI}{\affiliation{Space Telescope Science Institute, 3700 San Martin Drive, Baltimore, MD 21218-2410, USA}}
\newcommand{\UT}{\affiliation{University of Texas at Austin, 1 University Station C1400, Austin, TX 78712-0259, USA}}
\newcommand{\IoA}{\affiliation{Institute of Astronomy, University of Cambridge, Madingley Road, Cambridge CB3 0HA, UK}}
\newcommand{\QUB}{\affiliation{Astrophysics Research Centre, School of Mathematics and Physics, Queen's University Belfast, Belfast BT7 1NN, UK}}
\newcommand{\IPAC}{\affiliation{IPAC, Mail Code 100-22, Caltech, 1200 E. California Blvd., Pasadena, CA 91125, USA}}
\newcommand{\JPL}{\affiliation{Jet Propulsion Laboratory, California Institute of Technology, 4800 Oak Grove Dr, Pasadena, CA 91109, USA}}
\newcommand{\Southampton}{\affiliation{Department of Physics and Astronomy, University of Southampton, Southampton SO17 1BJ, UK}}
\newcommand{\LANL}{\affiliation{Space and Remote Sensing, MS B244, Los Alamos National Laboratory, Los Alamos, NM 87545, USA}}
\newcommand{\Tsinghua}{\affiliation{Physics Department and Tsinghua Center for Astrophysics, Tsinghua University, Beijing, 100084, People's Republic of China}}
\newcommand{\NAOC}{\affiliation{National Astronomical Observatory of China, Chinese Academy of Sciences, Beijing, 100012, People's Republic of China}}
\newcommand{\Itagaki}{\affiliation{Itagaki Astronomical Observatory, Yamagata 990-2492, Japan}}
\newcommand{\Einstein}{\altaffiliation{Einstein Fellow}}
\newcommand{\Hubble}{\altaffiliation{Hubble Fellow}}
\newcommand{\CfA}{\affiliation{Center for Astrophysics \textbar{} Harvard \& Smithsonian, 60 Garden Street, Cambridge, MA 02138-1516, USA}}
\newcommand{\UA}{\affiliation{Steward Observatory, University of Arizona, 933 North Cherry Avenue, Tucson, AZ 85721-0065, USA}}
\newcommand{\MPIA}{\affiliation{Max-Planck-Institut f\"ur Astrophysik, Karl-Schwarzschild-Stra\ss{}e 1, D-85748 Garching, Germany}}
\newcommand{\DSFP}{\altaffiliation{LSSTC Data Science Fellow}}
\newcommand{\HCO}{\affiliation{Harvard College Observatory, 60 Garden Street, Cambridge, MA 02138-1516, USA}}
\newcommand{\Carnegie}{\affiliation{Observatories of the Carnegie Institute for Science, 813 Santa Barbara Street, Pasadena, CA 91101-1232, USA}}
\newcommand{\TAU}{\affiliation{School of Physics and Astronomy, Tel Aviv University, Tel Aviv 69978, Israel}}
\newcommand{\Edinburgh}{\affiliation{Institute for Astronomy, University of Edinburgh, Royal Observatory, Blackford Hill EH9 3HJ, UK}}
\newcommand{\Birmingham}{\affiliation{Birmingham Institute for Gravitational Wave Astronomy and School of Physics and Astronomy, University of Birmingham, Birmingham B15 2TT, UK}}
\newcommand{\Bath}{\affiliation{Department of Physics, University of Bath, Claverton Down, Bath BA2 7AY, UK}}
\newcommand{\CTIO}{\affiliation{Cerro Tololo Inter-American Observatory, National Optical Astronomy Observatory, Casilla 603, La Serena, Chile}}
\newcommand{\Potsdam}{\affiliation{Institut f\"ur Physik und Astronomie, Universit\"at Potsdam, Haus 28, Karl-Liebknecht-Str. 24/25, D-14476 Potsdam-Golm, Germany}}
\newcommand{\INPE}{\affiliation{Instituto Nacional de Pesquisas Espaciais, Avenida dos Astronautas 1758, 12227-010, S\~ao Jos\'e dos Campos -- SP, Brazil}}
\newcommand{\UNC}{\affiliation{Department of Physics and Astronomy, University of North Carolina, 120 East Cameron Avenue, Chapel Hill, NC 27599, USA}}
\newcommand{\Ohio}{\affiliation{Astrophysical Institute, Department of Physics and Astronomy, 251B Clippinger Lab, Ohio University, Athens, OH 45701-2942, USA}}
\newcommand{\AAS}{\affiliation{American Astronomical Society, 1667 K~Street NW, Suite 800, Washington, DC 20006-1681, USA}}
\newcommand{\MMT}{\affiliation{MMT and Steward Observatories, University of Arizona, 933 North Cherry Avenue, Tucson, AZ 85721-0065, USA}}
\newcommand{\Geneva}{\affiliation{ISDC, Department of Astronomy, University of Geneva, Chemin d'\'Ecogia, 16 CH-1290 Versoix, Switzerland}}
\newcommand{\IUCAA}{\affiliation{Inter-University Center for Astronomy and Astrophysics, Post Bag 4, Ganeshkhind, Pune, Maharashtra 411007, India}}
\newcommand{\CMU}{\affiliation{Department of Physics, Carnegie Mellon University, 5000 Forbes Avenue, Pittsburgh, PA 15213-3815, USA}}
\newcommand{\NAOJ}{\affiliation{Division of Science, National Astronomical Observatory of Japan, 2-21-1 Osawa, Mitaka, Tokyo 181-8588, Japan}}
\newcommand{\IfA}{\affiliation{Institute for Astronomy, University of Hawai`i, 2680 Woodlawn Drive, Honolulu, HI 96822-1839, USA}}
\newcommand{\UCSC}{\affiliation{Department of Astronomy and Astrophysics, University of California, Santa Cruz, CA 95064-1077, USA}}
\newcommand{\Purdue}{\affiliation{Department of Physics and Astronomy, Purdue University, 525 Northwestern Avenue, West Lafayette, IN 47907-2036, USA}}
\newcommand{\Princeton}{\affiliation{Department of Astrophysical Sciences, Princeton University, 4 Ivy Lane, Princeton, NJ 08540-7219, USA}}
\newcommand{\Moore}{\affiliation{Gordon and Betty Moore Foundation, 1661 Page Mill Road, Palo Alto, CA 94304-1209, USA}}
\newcommand{\Durham}{\affiliation{Department of Physics, Durham University, South Road, Durham, DH1 3LE, UK}}
\newcommand{\JHU}{\affiliation{Department of Physics and Astronomy, The Johns Hopkins University, 3400 North Charles Street, Baltimore, MD 21218, USA}}
\newcommand{\Toronto}{\affiliation{David A.\ Dunlap Department of Astronomy and Astrophysics, University of Toronto,\\ 50 St.\ George Street, Toronto, Ontario, M5S 3H4 Canada}}
\newcommand{\Duke}{\affiliation{Department of Physics, Duke University, Campus Box 90305, Durham, NC 27708, USA}}
\newcommand{\NCU}{\affiliation{Graduate Institute of Astronomy, National Central University, 300 Jhongda Road, 32001 Jhongli, Taiwan}}
\newcommand{\Columbia}{\affiliation{Department of Physics and Columbia Astrophysics Laboratory, Columbia University, Pupin Hall, New York, NY 10027, USA}}
\newcommand{\Flatiron}{\affiliation{Center for Computational Astrophysics, Flatiron Institute, 162 5th Avenue, New York, NY 10010-5902, USA}}
\newcommand{\CIERA}{\affiliation{Center for Interdisciplinary Exploration and Research in Astrophysics (CIERA), 1800 Sherman Ave., Evanston, IL 60201, USA}}
\newcommand{\NU}{\affiliation{Department of Physics and Astronomy, Northwestern University, 2145 Sheridan Road, Evanston, IL 60208, USA}}
\newcommand{\SkAI}{\affiliation{NSF-Simons AI Institute for the Sky (SkAI), 172 E. Chestnut St., Chicago, IL 60611, USA}}
\newcommand{\GeminiNorth}{\affiliation{Gemini Observatory, 670 North A`ohoku Place, Hilo, HI 96720-2700, USA}}
\newcommand{\Keck}{\affiliation{W.~M.~Keck Observatory, 65-1120 M\=amalahoa Highway, Kamuela, HI 96743-8431, USA}}
\newcommand{\UW}{\affiliation{Department of Astronomy, University of Washington, 3910 15th Avenue NE, Seattle, WA 98195-0002, USA}}
\newcommand{\DiRAC}{\altaffiliation{DiRAC Fellow}}
\newcommand{\USask}{\affiliation{Department of Physics \& Engineering Physics, University of Saskatchewan, 116 Science Place, Saskatoon, SK S7N 5E2, Canada}}
\newcommand{\Thacher}{\affiliation{Thacher School, 5025 Thacher Road, Ojai, CA 93023-8304, USA}}
\newcommand{\Rutgers}{\affiliation{Department of Physics and Astronomy, Rutgers, the State University of New Jersey,\\136 Frelinghuysen Road, Piscataway, NJ 08854-8019, USA}}
\newcommand{\FSU}{\affiliation{Department of Physics, Florida State University, 77 Chieftan Way, Tallahassee, FL 32306-4350, USA}}
\newcommand{\Melbourne}{\affiliation{School of Physics, The University of Melbourne, Parkville, VIC 3010, Australia}}
\newcommand{\ASTROthreeD}{\affiliation{ARC Centre of Excellence for All Sky Astrophysics in 3 Dimensions (ASTRO 3D)}}
\newcommand{\Stromlo}{\affiliation{Mt.\ Stromlo Observatory, The Research School of Astronomy and Astrophysics, Australian National University, ACT 2601, Australia}}
\newcommand{\NCPAS}{\affiliation{National Centre for the Public Awareness of Science, Australian National University, Canberra, ACT 2611, Australia}}
\newcommand{\TAMU}{\affiliation{Department of Physics and Astronomy, Texas A\&M University, 4242 TAMU, College Station, TX 77843, USA}}
\newcommand{\Mitchell}{\affiliation{George P.\ and Cynthia Woods Mitchell Institute for Fundamental Physics \& Astronomy, College Station, TX 77843, USA}}
\newcommand{\ESO}{\affiliation{European Southern Observatory, Alonso de C\'ordova 3107, Casilla 19, Santiago, Chile}}
\newcommand{\MAS}{\affiliation{Millennium Institute of Astrophysics MAS, Nuncio Monsenor Sotero Sanz 100, Off.
104, Providencia, Santiago, Chile}}
\newcommand{\ICE}{\affiliation{Institute of Space Sciences (ICE, CSIC), Campus UAB, Carrer de Can Magrans, s/n, E-08193 Barcelona, Spain}}
\newcommand{\IEEC}{\affiliation{Institut d'Estudis Espacials de Catalunya, Gran Capit\`a, 2-4, Edifici Nexus, Desp.\ 201, E-08034 Barcelona, Spain}}
\newcommand{\Warwick}{\affiliation{Department of Physics, University of Warwick, Gibbet Hill Road, Coventry CV4 7AL, UK}}
\newcommand{\Macquarie}{\affiliation{School of Mathematical and Physical Sciences, Macquarie University, NSW 2109, Australia}}
\newcommand{\AAARC}{\affiliation{Astronomy, Astrophysics and Astrophotonics Research Centre, Macquarie University, Sydney, NSW 2109, Australia}}
\newcommand{\Capodimonte}{\affiliation{INAF - Capodimonte Astronomical Observatory, Salita Moiariello 16, I-80131 Napoli, Italy}}
\newcommand{\INFNNapoli}{\affiliation{INFN - Napoli, Strada Comunale Cinthia, I-80126 Napoli, Italy}}
\newcommand{\ICRANet}{\affiliation{ICRANet, Piazza della Repubblica 10, I-65122 Pescara, Italy}}
\newcommand{\MSU}{\affiliation{Center for Data Intensive and Time Domain Astronomy, Department of Physics and Astronomy,\\Michigan State University, East Lansing, MI 48824, USA}}
\newcommand{\SETI}{\affiliation{SETI Institute,
339 Bernardo Ave, Suite 200, Mountain View, CA 94043, USA}}
\newcommand{\IAIFI}{\affiliation{The NSF AI Institute for Artificial Intelligence and Fundamental Interactions}}
\newcommand{\ANUC}{\affiliation{Department of Astronomy, AlbaNova University Center, Stockholm University, SE-10691 Stockholm, Sweden}}
\newcommand{\UVA}{\affiliation{Department of Astronomy, University of Virginia, Charlottesville, VA 22904, USA}}
\newcommand{\THCA}{\affiliation{Physics Department and Tsinghua Center for Astrophysics (THCA), Tsinghua University, Beijing, 100084, People's Republic of China}}
\newcommand{\NARIT}{\affiliation{National Astronomical Research Institute of Thailand (NARIT), Don Kaeo, Mae Rim District, Chiang Mai 50180, Thailand}}
\newcommand{\HamObs}{\affiliation{Hamburger Sternwarte, Gojenbergsweg 112, 21029 Hamburg, Germany}}
\newcommand{\VaTech}{\affiliation{Department of Physics, Virginia Tech, 850 West Campus Drive, Blacksburg VA, 24061, USA}}
\newcommand{\Yunnan}{\affiliation{Yunnan Observatories, Chinese Academy of Sciences, Kunming 650216, P.R. China}}
\newcommand{\KLSECO}{\affiliation{Key Laboratory for the Structure and Evolution of Celestial Objects, Chinese Academy of Sciences, Kunming 650216, P.R. China}}
\newcommand{\ICS}{\affiliation{International Centre of Supernovae, Yunnan Key Laboratory, Kunming 650216, P.R. China}}
\newcommand{\FINCA}{\affiliation{Finnish Centre for Astronomy with ESO (FINCA), FI-20014 University of Turku, Finland}}
\newcommand{\Tuorla}{\affiliation{Tuorla Observatory, Department of Physics and Astronomy, FI-20014 University of Turku, Finland}}
\newcommand{\LAM}{\affiliation{Aix-Marseille Univ, CNRS, CNES, LAM, 13388 Marseille, France}}
\newcommand{\IAC}{\affiliation{Instituto de Astrof\'isica de Canarias, E-38205 La Laguna, Tenerife, Spain}}
\newcommand{\LPNHE}{\affiliation{LPNHE, (CNRS/IN2P3, Sorbonne Universit\'e, Universit\'e Paris Cit\'e), Laboratoire de Physique Nucl\'eaire et de Hautes \'Energies, 75005, Paris, France}}
\newcommand{\UniLag}{\affiliation{Universidad de La Laguna, Dept. Astrof\'isica, E-38206 La Laguna, Tenerife, Spain}}
\newcommand{\OU}{\affiliation{Homer L. Dodge Department of Physics and Astronomy, University of Oklahoma, 440 W. Brooks, Norman, OK 73019-2061, USA}}
\newcommand{\PSI}{\affiliation{Planetary Science Institute, 1700 East Fort Lowell Road, Suite 106, Tucson, AZ 85719-2395, USA}}
\newcommand{\TUM}{\affiliation{Technische Universit\"at M\"unchen, TUM School of Natural Sciences, Physik-Department, James-Franck-Stra\ss{}e 1, 85748 Garching, Germany}}
\newcommand{\UWarsaw}{\affiliation{Astronomical Observatory, University of Warsaw, Al. Ujazdowskie 4, 00-478 Warszawa, Poland}}
\newcommand{\Trinity}{\affiliation{School of Physics, Trinity College Dublin, The University of Dublin, Dublin
2, Ireland}}
\newcommand{\UNAB}{\affiliation{Departamento de Ciencias F\'isicas, Facultad de Ciencias Exactas, Universidad Andr\'es Bello, Fern\'andez Concha 700, Las Condes,
Santiago, Chile}}
\newcommand{\CCAPP}{\affiliation{Center for Cosmology and Astroparticle Physics, The Ohio State University, 191 West Woodruff Ave, Columbus, OH, USA}}
\newcommand{\OSU}{\affiliation{Department of Astronomy, The Ohio State University, 140 West 18th Avenue, Columbus, OH, USA}}
\newcommand{\LCOactual}{\affiliation{Las Campanas Observatory, Carnegie Observatories, Casilla 601, La
Serena, Chile}}
\newcommand{\UCSD}{\affiliation{Department of Astronomy \& Astrophysics, University of California, San Diego, 9500 Gilman Drive, MC 0424, La Jolla, CA 92093-0424, USA}}
\newcommand{\Liverpool}{\affiliation{Astrophysics Research Institute, Liverpool John Moores University, 146 Brownlow Hill, Liverpool, L3 5RF, UK}}

\newcommand{\Monash}{\affiliation{School of Physics and Astronomy, Monash University, Clayton, Victoria 3800, Australia}}
\newcommand{\OzGrav}{\affiliation{OzGrav: The ARC Centre of Excellence for Gravitational Wave Discovery, Clayton, Victoria 3800, Australia}}
\author[0000-0001-8073-8731]{Bhagya~M.~Subrayan}
\UA
\email[show]{bsubrayan@arizona.edu}
\author[0000-0003-4102-380X]{David J. Sand}
\UA
\email{dsand@arizona.edu}

\author[0009-0002-5272-1929]{Olivia Culbert}
\UA
\email{oliviaculbert@arizona.edu}

\author[0000-0003-0123-0062]{Jennifer E. Andrews}
\GeminiNorth
\email{jennifer.andrews@noirlab.edu}

\author[0000-0002-0744-0047]{Jeniveve Pearson}
\UA
\email{jenivevepearson@arizona.edu}

\author[0000-0002-0832-2974]{Griffin Hosseinzadeh}
\UCSD
\email{ghosseinzadeh@ucsd.edu}

\author[0000-0001-8738-6011]{Saurabh W.\ Jha}
\Rutgers
\email{saurabh@physics.rutgers.edu}

\author[0000-0001-8818-0795]{Stefano Valenti}
\UCD
\email{stfn.valenti@gmail.com}
\author[0000-0002-4924-444X]{K. Azalee Bostroem}
\altaffiliation{LSST-DA Catalyst Fellow}
\IPAC
\email{bostroem@ipac.caltech.edu}

\author[0000-0003-4175-4960]{Conor~L.~Ransome}
\UA
\email{cransome@arizona.edu} 

\author[0000-0002-7352-7845]{Aravind P.\ Ravi}
\UCD
\email{apazhayathravi@ucdavis.edu}
\author[0000-0002-9085-8187]{Aysha Aamer}
\UA
\email{aamer2@arizona.edu}

\author[0000-0002-1895-6639
]{Moira Andrews}
\LCO\UCSB
\email{mandrews@lco.global}


\author[0000-0003-4666-4606]{Emma R. Beasor}
\Liverpool
\email{E.R.Beasor@ljmu.ac.uk}

\author[0000-0003-0528-202X]{Collin Christy}
\UA 
\email{collinchristy@arizona.edu}

\author[0000-0002-7937-6371]{Yize Dong \begin{CJK*}{UTF8}{gbsn}(董一泽)\end{CJK*}}
\CfA
\email{yizdong@ucdavis.edu}

\author[0000-0003-4537-3575]{Noah Franz}
\UA 
\email{nfranz@arizona.edu}

\author[0000-0003-2744-4755]{Emily Hoang}
\UCD
\email{emthoang@ucdavis.edu}

\author[0000-0002-9454-1742]{Brian Hsu}
\UA
\email{bhsu@arizona.edu}

\author[0000-0001-5754-4007]{Jacob Jencson}
\IPAC
\email{jjencson@ipac.caltech.edu}

\author[0000-0003-3108-1328]{Lindsey A. Kwok}
\CIERA
\email{lindsey.kwok@northwestern.edu}

\author[0000-0001-9589-3793]{M.~J. Lundquist}
\Keck
\email{mlundquist@keck.hawaii.edu}

\author[0009-0008-9693-4348]{Darshana Mehta}
\UCD
\email{ddmehta@ucdavis.edu}

\author[0000-0002-7015-3446]{Nicol\'as Meza Retamal}
\UCD
\email{}

\author[0000-0002-4022-1874]{Manisha Shrestha}
\Monash\OzGrav
\email{Manisha.Shrestha@monash.edu}

\author[0000-0001-5510-2424]{Nathan Smith}
\UA
\email{nathans@as.arizona.edu}
\author[0000-0002-4951-8762]{Sergiy Vasylyev}
\UCSD
\email{svasylyev@ucsd.edu}

\begin{abstract}

Core collapse supernovae (CCSNe) are  invoked as major dust producers in the early Universe, yet the amount of dust they form, the timescale over which it grows, and the physical conditions that regulate the yield remain uncertain. We present a detailed JWST/MIRI mid-infrared (MIR) imaging census of 11 nearby Type IIP  CCSNe spanning $\sim$1--7 yr after explosion, investigating dust emission across the different phases of their evolution. The spectral energy distributions show a coherent evolution from hot ($\sim$1500 K), 5–8 $\mu$m emission at $\sim$400 d to prominent 10 and 18 $\mu$m emission features at later epochs ($\sim$600–2500 d). The cool dust temperatures range from $\sim120$ to $250$~K and dust masses from $\sim10^{-4}$ to $10^{-2} M_{\odot}$. The current sample shows no statistically significant correlation between measured dust mass and peak luminosity, plateau duration, or explosion energy. SNe with early high ionization features, indicative of confined CSM, are often among the dust-rich objects in the sample. The measured 1-7 yr dust yields are insufficient to account for dust in typical $z >6$ galaxies, but support a role for CCSNe as producers of seed dust for subsequent grain growth.

\end{abstract}


\section{Introduction}\label{sec:intro}

Cosmic dust is a cornerstone of galaxy evolution, driving the physics and chemistry of the interstellar medium (ISM), regulating star formation, and reprocessing roughly half of all starlight in the Universe \citep{Hollenbach1979,Dwek1987, Granato2000, Yamasawa2011, Dian2020}. Yet the origin of this dust remains uncertain. Observations with the Atacama Large Millimeter/submillimeter Array (ALMA) and the James Webb Space Telescope (JWST) have revealed that substantial dust reservoirs, exceeding $\sim 10^6-10^8\,M_\odot$, were already in place in galaxies less than 1~Gyr after the Big Bang \citep[e.g.,][]{Marrone18, Hashimoto19, Witstok23, Nanni2025}. The long evolutionary timescales of asymptotic giant branch (AGB) stars, likely the dominant dust factories in the local Universe, are inconsistent with the rapid dust enrichment observed at early epochs. This mismatch has shifted attention to core-collapse supernovae (CCSNe) as the primary candidate dust producers in the early Universe, since their short lifetimes allow them to rapidly seed it with dust \citep{Morgan03_earlygal, Maiolino04, Gall11, Schneider24}.

To satisfy the high-redshift dust budgets, models require that each CCSN produce between $0.1$ and $1\,M_\odot$ of dust \citep{Todini01, Sarangi18, Schneider24}. Far-infrared and sub-millimeter observations of young supernova remnants including SN 1987A, Cassiopeia A (Cas A), and the Crab Nebula have revealed substantial reservoirs of cold ejecta dust, with masses of $\sim 0.1-0.8\,M_\odot$  building up on timescales of decades to centuries after the explosion \citep{Dunne03, Morgan03_keplers, Matsuura11, Matsuura2015, Wessen2015_SN1987A, Dwek2015, Cigan2019, Chastenet22, Priestley22, Milisavljevic2024, Rho2024_CasA, Temim2024_Crab}. However, the vast majority of pre-JWST near- and mid-infrared (MIR) studies of extragalactic CCSNe in the first few years to decades after explosion have inferred dust masses that are orders of magnitude lower, typically $\lesssim10^{-2}\,M_\odot$ \citep{Gall11, Andrews2011_2007it, Szalai13, Szalai19_spitzer,Fox2011,Tinyanont16}. This is in large part a consequence of the limited wavelength coverage available prior to JWST. After its cryogen was exhausted in May 2009, Spitzer's warm extended mission (2009--2020) was restricted to only the 3.6 and 4.5~$\mu$m IRAC channels \citep{Fazio2004, Werner2004}, confining photometric sensitivity to warm dust ($\gtrsim$500~K) and leaving cool grain populations peaking at $\gtrsim$10~$\mu$m largely inaccessible to extragalactic SN observations for over a decade. Thus, traditional IR observations so far may have missed a significant fraction of newly formed dust, either because it is too cold to emit at these wavelengths or because it is hidden behind large optical depths in the clumpy ejecta \citep{Wessen2015_SN1987A,Bevan2016_1987A, Niculescu-Duvaz2022}.

Dust emission in CCSNe can arise from multiple physical mechanisms, but single-epoch broadband photometry primarily constrains the total dust mass and temperature rather than its origin. Pre-existing dust in the progenitor's circumstellar environment, shaped by winds or eruptive mass loss prior to core collapse, can produce infrared echoes when heated by the SN flash or by ongoing shock interaction, dependent on the mass-loss history and dust survival radius of the progenitor system \citep{Bode80,Dwek83, Fox10, Fox2011, Hosseinzadeh23}. Newly formed dust can instead arise in two principal sites. In the freely expanding metal-rich ejecta, grain nucleation begins once the gas cools sufficiently, typically below $\sim2000$ K, with warm dust masses of order $10^{-4}$--$10^{-3} M_\odot$ building up over $\sim$ 400 - 600 days post-explosion \citep{Wooden1993,Kotak09, Fabbri2011_SN2004et, Sarangi_2013, Sarangi18}. In SNe with strong ejecta-circumstellar material (hereafter CSM) interaction \citep{Chevalier1994}, dust can also form in the cold dense shell (hereafter CDS) between the forward and reverse shocks, where high post-shock densities and efficient radiative cooling can create favorable conditions for molecule and grain formation \citep{Pozzo04, Mattila08, Smith2008_SN2006jc, Smith09}. Theoretical models predict that the interaction powered channel may accelerate dust formation relative to freely expanding ejecta, although the inferred dust mass and formation timescale depend sensitively on the density, geometry, and strength of the CSM interaction \citep{Sarangi_2013, Sarangi_Slavin2022, Takei2025}.

The signatures of dust condensation in CCSNe are diagnosed through blueshifted, red-attenuated emission-line profiles from selective extinction of the receding ejecta \citep{Lucy89,Weil2020,Shahbandeh23,Singh2026_23ixf}, optical fading beyond the $^{56}$Co decay rate as newly formed grains absorb radioactive luminosity, and a rising thermal infrared excess from reprocessed emission \citep{Wooden1993}. Carbon monoxide (CO) emission typically precedes these signatures and thus serves as an early molecular
tracer of the conditions required for dust formation \citep{Rho18,Tinyanont19,Sarangi18,Park2025_SN2023ixf,Medler2026,Mera2026_SN2024ggi}. Although optical fading and IR brightening are often interpreted as signatures of dust formation, the same behavior can also arise if radiation from a declining ejecta-CSM interaction is reprocessed by pre-existing CSM dust; in that case, line profile asymmetries provide a more direct diagnostic of dust co-spatial with the emitting ejecta \citep{Fox10, Andrews2010_SN2007od, Shahbandeh24}.
At later epochs, the reverse shock sweeping back through the ejecta can destroy an uncertain fraction of the newly condensed dust, leaving net grain survival one of the key unknowns in quantifying the dust yield of CCSNe \citep{BS2007, Nozawa2007, Silvia2010, Kirchschlager2019, Priestley22}. 

JWST has opened a new window for characterizing dust reservoirs in nearby extragalactic CCSNe. Existing late-time studies have largely focused on objects pre-selected for their known infrared brightness, prior Spitzer detections, or confirmed CSM interaction (e.g. Type IIn) \citep{Arendt2023_SN1987A, Jones2023_SN1987A, Shahbandeh23, Shahbandeh24, Zsiros24, Clayton2025_SN1995N, Sarangi2025_SN2005af, Szalai2025_SN1993J, Tinyanont2025_SN2014C, Pearson2025,Davis2026_SN2023xgo}. Meanwhile, phase resolved panchromatic campaigns tracing molecule and dust formation from the nebular epoch onward remain limited to the exceptionally nearby SNe~2023ixf and 2024ggi \citep{Derkacy2026_23ixf, Singh_2024_SN2023ixf_asphericity,Medler2025_SN2023ixf, Baron2025, Mera2026_SN2024ggi}. What has been missing is a phase-resolved JWST survey of a representative sample of ``normal" Type~IIP SNe, the dominant fraction of hydrogen-rich CCSNe \citep{Li2011, Smith2011} to determine whether the dust masses uncovered so far are typical of CCSNe or a property of an IR-selected, dust-bright subset.

\begin{table*}[ht]
\centering
\setlength{\tabcolsep}{2pt}

\begin{tabular}{lllcccccc}
\hline
 & \textbf{SN} & \textbf{RA} & \textbf{Dec} & \textbf{$\approx$ Dist} & \textbf{Explosion Epoch} & \textbf{Phase} & \textbf{E(B-V)} & \textbf{Reference} \\
 &  &  & & \textbf{(Mpc)} & \textbf{(yyyy-mm-dd)} & \textbf{(days)}  & \textbf{mag}  &  \\
\hline 
\multirow{3}{*}{\rotatebox{90}{$\sim$2 yr}}  & SN~2022wsp & 23:00:03.53 & +15:58:42.60 & 26 & 2022, Oct 2 &  401 & $0.35 \pm 0.10 $ & \citet{SN2022wsp_Vasylev} \\
& SN~2022acko & 03:19:38.99 & $-$19:23:42.68 & 19 & 2022, Dec 5 & 405 & $0.05 \pm 0.01$ & \citet{22acko_Bostroem}  \\
& SN~2024ggi & 11:18:22.09 & $-$32:50:15.3 & 7 & 2024, Apr 11 & 660 & $0.15 \pm 0.04$ & \citet{Shrestha2024}  \\
 & SN~2022jox & 09:57:44.52 & $-$28:30:57.06 & 38 & 2022, May 9 & 701 & $0.09 \pm 0.01 $  & \citet{22jox_Andrews2024}  \\
\hline
\multirow{3}{*}{\rotatebox{90}{$\sim$ 3 yr}} & SN~2021yja & 03:24:21.18 & $-$21:33:56.20 & 25 & 2021, Sept 7 & 859 & $0.10 \pm 0.02$  & \citet{Hosseinzadeh2022} \\
& SN~2023ixf & 14:03:38.56 & +54:18:42.0 & 7 & 2023, May 19 & 992 & $0.03 \pm 0.01$  & \citet{Hosseinzadeh_2023_23ixf} \\
& SN~2021gmj & 10:38:47.27 & +53:30:30.31 & 18 & 2021, Mar 19 & 1120 & $0.05 \pm 0.01$ &  \citet{21gmj_Meza_Retamal_2024}\\
& SN~2020jfo & 12:21:50.48 & +04:28:54.05 & 15 & 2020, May 5 & 1471 & $0.29 \pm 0.05$  & \citet{20jfo_Ailawadhi} \\
\hline
\multirow{3}{*}{\rotatebox{90}{ 6-7 yr}} & SN~2018cuf & 21:16:11.55 & $-$64:28:57.50 & 42 & 2018, Jun 23  & 1929  & $0.14 \pm 0.01$ & \citet{2018cuf_Yize} \\
& SN~2017eaw & 20:34:44.24 & +60:11:35.84 & 7 & 2017, May 14& 2330 & $0.34  \pm 0.05$ & \citet{Kilpatrick18} \\
& SN~2017gmr & 02:35:30.18 & $-$09:21:15.08 & 20 & 2017, Sept 3 & 2335 & $0.30  \pm 0.10$ & \citet{17gmr_Andrews_Jen} \\

\hline
\end{tabular}
\caption{ Properties of the CCSN sample ordered by phase of the JWST/MIRI observations in rest frame days after explosion The table lists the coordinates, host galaxy distances, explosion epochs, as well as total extinction grouped into recent, intermediate-age, and older events.} \label{tab:sample}
\end{table*}

Here we present results from a JWST/MIRI observational campaign designed to address this gap. We assembled a sample of 11 nearby Type IIP SNe spanning three post-explosion phases for dust formation and evolution, from $\sim$400 to $\sim$2300 days. The sample includes nine CCSNe, observed in JWST Cycle 2, drawn from the normal Type IIP population \citep{Anderson2014, Valenti2016, Gutirrez2017, Martinez2022}, selected without regard to prior infrared brightness or CSM signatures. We additionally include two well studied CSM-interacting SNe, SN 2023ixf and SN 2024ggi, observed in JWST Cycle 4 for comparison. Combining JWST/MIRI photometry with a comprehensive literature compilation, we trace the dust formation timeline in normal SNe~IIP and investigate the roles of intrinsic SN properties and the circumstellar environment in shaping the observed dust mass. The paper is structured as follows. Section~\ref{sec:observations} describes our sample selection, observations, and data reduction. Section \ref{sec:seds} describes the JWST/MIRI SEDs of the overall sample, followed by Section~\ref{sec:modeling} that details the dust SED modeling methods. Section~\ref{sec:results_discussions} presents the derived dust properties for each supernova and discusses sample wide trends.  We also compare dust properties with a larger sample across CCSN subtypes and discuss the implications for SN dust production. Finally, Section~\ref{sec:conclusion} summarizes our main conclusions.

\section{Observations}\label{sec:observations}

\subsection{Sample Selection}

\begin{figure}[htp]
	\centering
\includegraphics[width = \columnwidth]{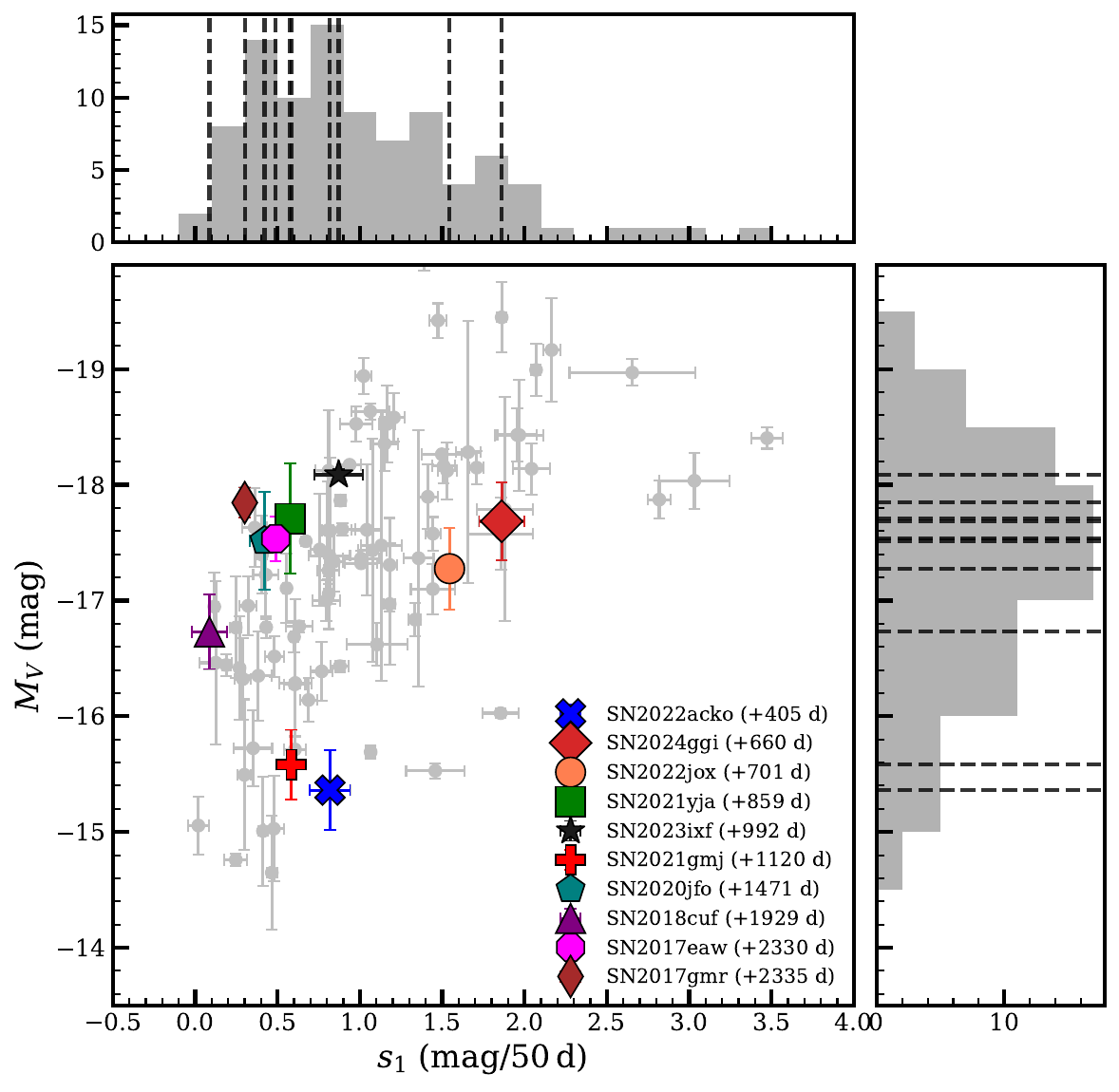}
    \caption{Peak absolute $V$-band magnitude, $M_V$, compared to the early decline slope, $s_1$, for our JWST sample (colored symbols) and a comparison sample of SNe~II drawn from \citet{Anderson2014, Valenti2016,deJaeger2019,Anderson2024}.}
    \label{fig:SNII_Sample}
\end{figure}

We selected nine normal Type~IIP SNe spanning three post-explosion phase bins, $\sim$1--2 yr, $\sim$2--5 yr, and $\gtrsim$5 yr, with three objects per bin to sample the expected $\sim$1--7 yr dust-growth timescale \citep{Sarangi_Chercheneff2015}. The sample covers a range of peak luminosities, plateau durations, and early CSM signatures, including narrow flash-ionization features, enabling an initial assessment of how dust properties relate to SN diversity. All targets were discovered within days of explosion and have well-sampled multi-band photometry; preference for objects with strong \textit{HST} legacy data naturally selects distances of $\sim$10--40 Mpc. This sample was not selected based on prior dust emission. SN~2017eaw, which has known dust signatures and prior Cycle~1 JWST observations \citep[PID~2666;][]{Shahbandeh23} is still included in the sample as a well-studied nearby CCSN. We also observed the known CSM-interacting SNe~2023ixf and 2024ggi in Cycle~4 as comparison objects to evaluate how enhanced interaction modifies dust formation relative to the normal Type~II population. Figure~\ref{fig:SNII_Sample} places the sample in the peak
luminosity-decline rate ($M_V$-$s_1$) parameter space against the representative Type~II population.The properties of the sample, including coordinates, distances, explosion epochs, phases, and extinction, are provided in Table~\ref{tab:sample}, grouped by approximate age since explosion.

\begin{figure*}[htp]
	\centering
\includegraphics[width = \textwidth]{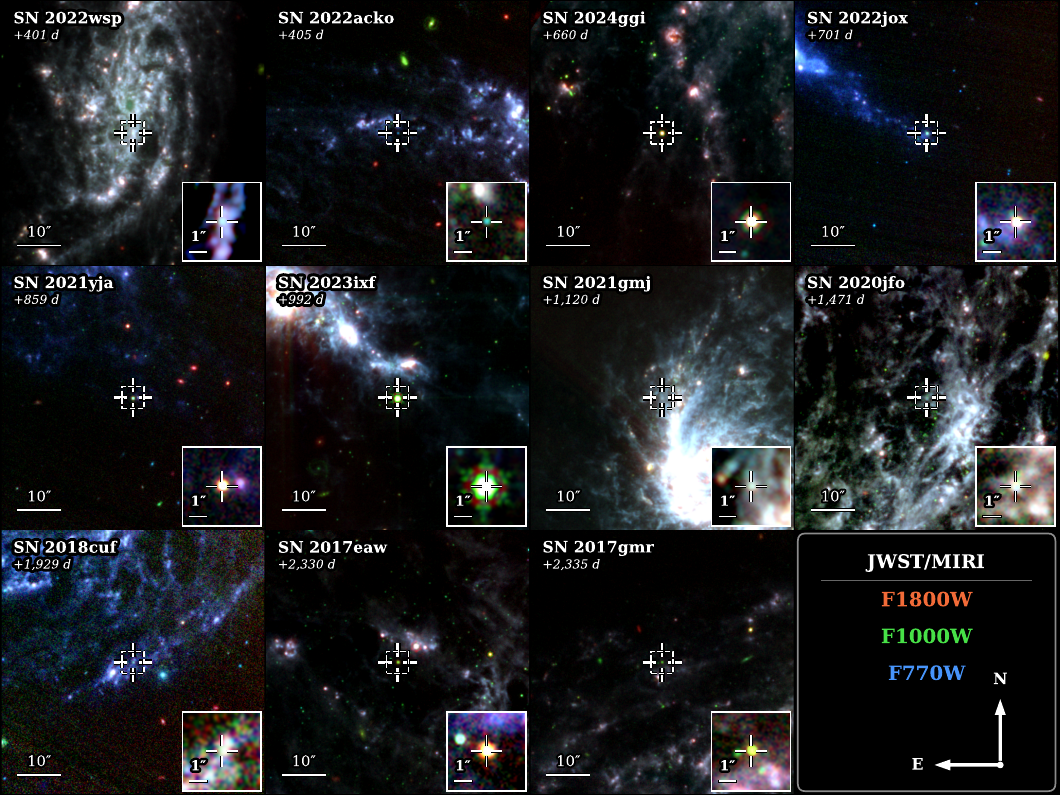}
   \caption{ JWST/MIRI three-color images of the SN sample, constructed from F770W (blue), F1000W (green), and F1800W (red). Each panel is centered on the SN position, marked by white cross hairs, and labeled with phase relative to explosion. Several SNe lie in complex host environments, including spiral arms, star-forming regions, and structured diffuse MIR backgrounds. Insets show zoomed views of the SN locations. 
\label{fig:miri_rgb_grid}
}
    \label{fig:pretty_picture}
\end{figure*}

\subsection{JWST MIRI}

JWST Mid-Infrared Instrument (MIRI; \citealt{Bouchet2015,Rieke2015,Wright2023}) observations were obtained as part of JWST Cycle~2 GO~3295 \citep{Cycle2JWST} and Cycle~4 GO~7881 \citep{Cycle4JWST}. These programs provide late-time MIR imaging over phases of approximately 400--2335 days after explosion, with Cycle~2 observations of the normal Type~II sample obtained between October 2023 and May 2024 and Cycle~4 observations targeting SN~2023ixf and SN~2024ggi obtained in February 2026. All observations used the full MIRI imaging filter set, F560W, F770W, F1000W, F1130W, F1280W, F1500W, F1800W, F2100W, and F2550W, with the FULL array, FASTR1 readout pattern, a four-point dither sequence, and an exposure time of 111~s per filter. Point-spread-function (PSF) photometry was performed using \texttt{space\textunderscore phot} on the Level~2 calibrated products for all filters except F2550W \citep{pierel_space_phot}. For all other filters, the SN PSF was independently fit in each of the four Level~2 CAL files using WebbPSF models \citep[version~1.2.1;][]{WebbPSF1, WebbPSF2}. Owing to the low signal-to-noise ratio in F2550W, PSF photometry for this band was instead carried out on the Level~3 stacked image. In this case, \texttt{space\textunderscore phot} constructs a Level~3 PSF model by drizzling temporally and spatially dependent Level~2 WebbPSF models, with reported uncertainties that likely underestimate the true photometric errors, likely for F2550W. For non-detections, we report $3\sigma$ flux upper limits, with $\sigma$ estimated from the root mean square of background fluxes measured using apertures matched to the source aperture. We note that the majority of the measurements in our sample are significant detections.

For high signal-to-noise detections at shorter wavelengths, the local background was estimated from two nearby source-free regions on either side of the target aperture and subtracted before flux measurement, with aperture corrections derived from WebbPSF models following \citet{Pearson2025}. At longer wavelengths, where the MIRI PSF is broader and the measurements are more sensitive to instrumental and complex host galaxy background, the background was estimated from nearby regions chosen to sample the local environment while avoiding the strongest visible structures. The estimated background level and its uncertainty were propagated into the final photometric errors, following the methodology described in \citet{Hosseinzadeh23}. The resulting flux measurements from \texttt{space\textunderscore phot} for the sample are presented in Table~\ref{tab:miri_photometry}. Figure~\ref{fig:miri_rgb_grid} shows the JWST/MIRI F770W--F1000W--F1800W color composites for the full sample.

\subsection{MMT NIR Imaging}

We obtained ground-based near-infrared (NIR) photometry for SN~2020jfo, SN~2021gmj, SN~2022wsp, and SN~2023ixf using the MMT and Magellan Infrared Spectrograph (MMIRS) on the MMT \citep{MMIRS}. Observations were conducted in the $JHK$ bands during June 2025 (SN~2020jfo), January 2025 (SN~2021gmj), November 2025 (SN~2022wsp) and February 2026 (SN~2023ixf). Each MMIRS observation comprised a dithered sequence alternating between the supernova field and an adjacent off-galaxy sky field to facilitate accurate sky subtraction, particularly given the high surface brightness of the host galaxies in the NIR. The MMIRS data were reduced using POTPyRI \footnote{Adapted from the publicly available MMIRS imaging pipeline: \url{https://github.com/CIERA-Transients/POTPyRI}} that performs standard dark-current subtraction, flat-field correction, sky background estimation and subtraction, astrometric alignment, and stacking of the individual exposures. The MMIRS field of view (6\farcm9~$\times$~6\farcm9) is sufficiently large to enable photometric calibration using isolated stars with cataloged magnitudes from the Two Micron All Sky Survey \citep[2MASS;][]{2MASS}. For each image, we derived an effective PSF (ePSF) model by fitting bright, isolated stars with the \texttt{EPSFBuilder} tool from the \texttt{photutils} package in \texttt{Astropy}. PSF-fitting photometry was then performed at the supernova position and for a set of 20 or more well-distributed field stars. To account for spatial variations in the background, we included a low-order, two-dimensional polynomial term in the PSF model. The statistical uncertainty of each flux measurement was estimated per pixel using the root-mean-square (RMS) of the fitting residuals, scaled by $\sqrt{\chi^2_\nu}$ (typically $\gtrsim 1$), and multiplied by the number of noise pixels in the ePSF.\footnote{Noise pixel definition following F. Masci: \url{http://web.ipac.caltech.edu/staff/fmasci/home/mystats/noisepix_specs.pdf}}

Photometric zeropoints and aperture corrections (typically $\lesssim 0.1$~mag) were derived from the calibration stars and applied to scale the PSF-fitting magnitudes. The total photometric uncertainty was obtained by summing in quadrature the statistical flux uncertainty, the RMS dispersion of the zero-point calibration stars, and the uncertainty in the ePSF aperture correction. The zero-point uncertainty was estimated as \(\mathrm{RMS}/\sqrt{N}\), where \(N\) is the number of equally weighted calibration stars. Table \ref{tab:nir_fluxes} summarizes the NIR photometry for the sample. For SN 2020jfo, SN 2021gmj, and SN 2022wsp, we do not detect significant NIR emission associated with the SN itself so we treat them as upper limits.

\subsection{Very Late-Time Optical Spectroscopy}

To complement the JWST Cycle~2/Cycle~4 MIRI observations, we conducted a very late-time optical spectroscopic campaign targeting the full supernova sample using Keck~I and Gemini North/South. The Gemini observations were obtained under programs GN-2024A-Q-139/GS-2024A-Q-137 (PI: Andrews) and GN-2026A-Q-124/GS-2026A-Q-120 (PI: Subrayan), while the Keck observations were carried out under NASA-Keck program N017 (PI: Sand) and UC Davis program time (PI: Valenti). All eleven SNe were targeted between phases corresponding to $\sim$ 600-2600~days post-explosion, with several objects observed at multiple epochs, however, only five were detected with sufficient signal-to-noise for spectroscopic analysis. The Keck observations were obtained with the Low Resolution Imaging Spectrometer (LRIS; \citealt{Oke95}) using a $1\arcsec$ slit, the 560~nm dichroic, the 600/4000 grism on the blue arm, and the 400/8500 grating on the red arm. The Gemini observations were obtained with the Gemini Multi-Object Spectrograph (GMOS; \citealt{Hook2004,Gimeno2016}) using a $1\arcsec$ slit and the B600 grating centered at 680~nm. The LRIS and GMOS data were reduced with \texttt{LPipe} \citep{Perley2019} and \texttt{DRAGONS} \citep{Labrie2019}, respectively, following standard long-slit reduction procedures, including bias correction, flat-fielding, wavelength calibration, sky subtraction, and flux calibration. Table~\ref{tab:spec_log} summarizes the observing log for observations with clear traces of the SN. The SN~2017eaw spectrum used in this work is adopted from \citet{Pearson2025}.
\begin{deluxetable}{lccc}
\label{tab:spec_log}
\tablecaption{Log of Spectroscopic Observations}
\tablehead{
\colhead{SN} & \colhead{MJD} & \colhead{Telescope/Instrument} & \colhead{Phase (d)}
}
\startdata
SN~2022acko & 60556.56 & Keck/LRIS & 638 \\
SN~2024ggi  & 61082.27 & Gemini-S/GMOS & 671 \\
SN~2023ixf  & 61084.59 & Gemini-N/GMOS & 1002 \\
SN~2020jfo  & 60350.30 & Gemini-S/GMOS & 1376 \\
SN~2017eaw  & 60553.25 & Keck/LRIS & 2668
\enddata
\end{deluxetable}

\section{JWST/MIRI SED\lowercase{s} of the Sample}\label{sec:seds}

\begin{figure*}[htp]
	\centering
\includegraphics[width =0.95\textwidth]{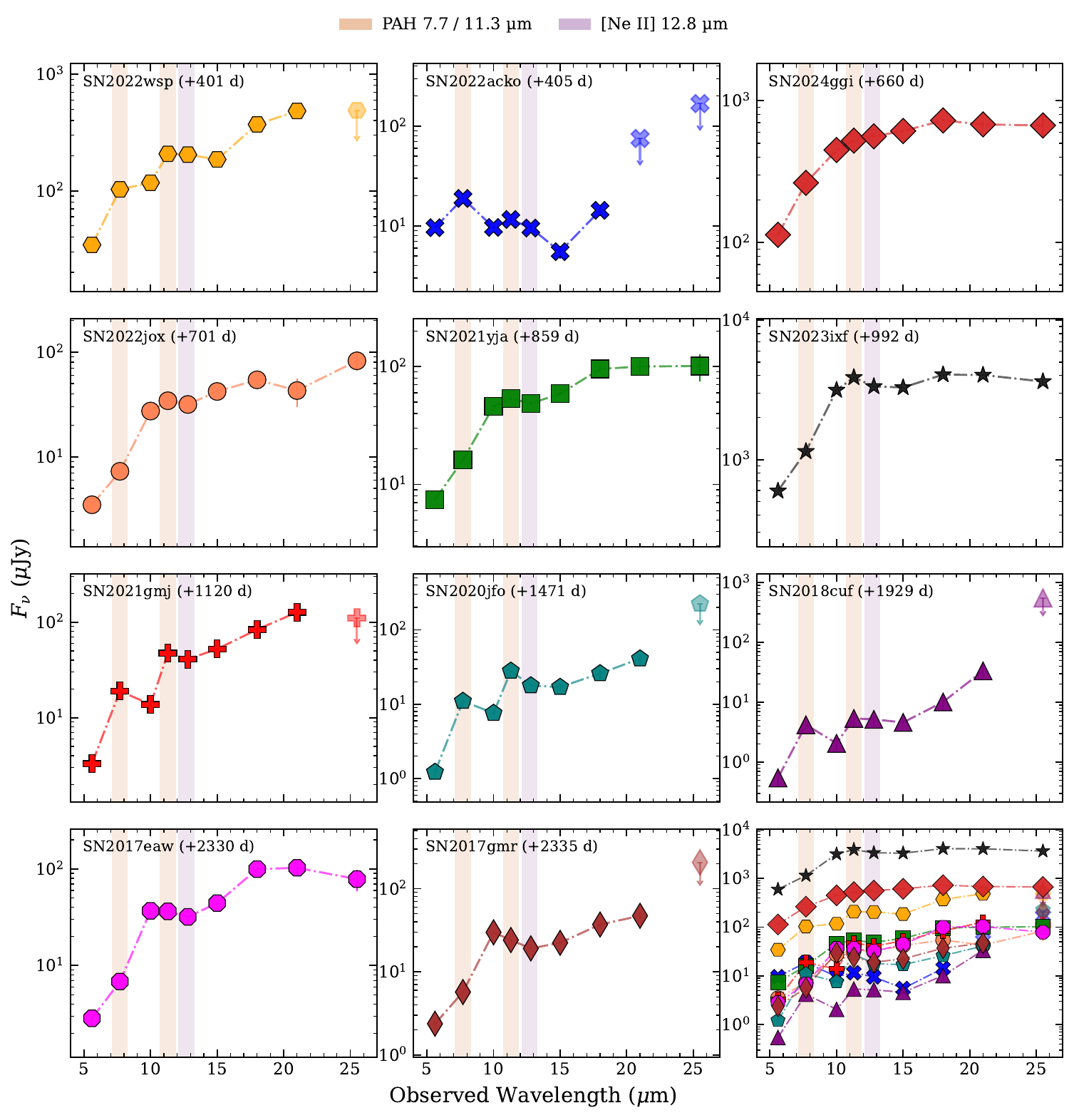}
 \vspace{-5pt}
    \caption{JWST/MIRI spectral energy distributions for nine ``normal" Type~II SNe spanning  $+401$ to $+2335$~d post-explosion. Flux densities ($F_\nu$) are plotted against wavelength on a logarithmic scale. Downward-pointing arrows indicate $3\sigma$ upper limits for non-detections. The shaded regions indicate the F770W, F1130W and F1280W filter bandpasses, which coincide with the PAH and [\ion{Ne}{2}] emission features, respectively. The last panel shows the overall comparison of the sample SEDs. }
    \label{fig:sed_all}
\end{figure*}

The JWST/MIRI SEDs of the full sample are shown in Figure~\ref{fig:sed_all}, presenting MIR flux densities from JWST/MIRI spanning $\sim$5--25~$\mu$m. 
In this section, we present a qualitative assessment of the SED morphology across the sample; a quantitative dust modeling analysis is presented in Section~\ref{sec:modeling}. 

The sample spans a broad range of post-explosion phases, from the youngest objects, SN~2022wsp and SN~2022acko, at $\sim$400~days, to the oldest events, SN~2017eaw and SN~2017gmr, at $\gtrsim$2300~days. All 11 SNe are detected in at least seven MIRI filters. Flux densities span more than three orders of magnitude, from $\sim$0.55~$\mu$Jy (SN~2018cuf at F560W) to $\sim$3880~$\mu$Jy (SN~2023ixf at F1130W). SN~2018cuf, SN~2020jfo, and SN~2022acko fall on the lower end of the flux densities as compared to SN~2023ixf, SN~2024ggi, and SN~2022wsp. 

Across the full sample, all 11 objects exhibit a rising MIR continuum from $\sim$5~$\mu$m toward longer wavelengths, broadly consistent with thermal dust emission \citep[e.g.,][]{Gall11,Szalai19, Shahbandeh23}. At the earliest phases ($\sim$400--660~days), the SEDs likely include contributions from both a residual optically thick photospheric component and emerging dust emission, with silicate dust expected to dominate the IR emission in O-rich environments, whether newly formed in the ejecta or pre-existing in the CSM \citep{Todini01,Nozawa2003, Sarangi18, Sarangi2022, Sarangi2025_SN2005af}. The $\sim$10~$\mu$m feature is attributed to Si--O stretching modes in silicates, while the $\sim$18~$\mu$m feature is attributed to O--Si--O bending modes in the same material \citep{Dorschner1995,Henning2010, Laor93}. At comparable early epochs, however, the SEDs show clear differences in both luminosity and shape. SN~2022wsp and SN~2022acko are both at $\sim$400~days, yet SN~2022wsp rises smoothly across the MIRI wavelength range, while SN~2022acko displays a non-monotonic SED. The $\sim$10~$\mu$m silicate feature is generally weak in both objects at this phase; however, SN~2022wsp shows rising emission components near $\sim$18~$\mu$m, whereas SN~2022acko, owing to its faintness, has only upper limits in the reddest filters. 

\begin{deluxetable}{l l c c c}[t]
\label{tab:nir_fluxes}
\tablecaption{Late-time NIR imaging observations obtained with MMT/MMIRS}
\tablehead{
\colhead{SN} & \colhead{Filter} & \colhead{MJD} & \colhead{Phase [d]} & \colhead{Vega Mag}
}
\startdata
SN~2023ixf & J & 61075.39 & 992  & $19.74 \pm 0.17$ \\
           & H & 61075.39 & 992  & $18.38 \pm 0.21$ \\
           & K & 61075.40 & 992  & $16.96 \pm 0.24$ \\
SN~2020jfo & J & 60841.18 & 1867 & $> 20.33$ \\
           & H & 60842.20 & 1868 & $> 19.86$ \\
SN~2021gmj & J & 60690.40 & 1398 & $> 17.26$ \\
           & J & 60694.34 & 1402 & $> 17.17$ \\
           & H & 60690.35 & 1398 & $> 16.58$ \\
           & H & 60694.35 & 1402 & $> 16.66$ \\
           & K & 60690.38 & 1398 & $> 16.23$ \\
           & K & 60694.35 & 1402 & $> 16.26$ \\
SN~2022wsp & J & 60867.457 & 1013 & $> 15.51$ 
\enddata
\end{deluxetable}

Among the SNe observed at phases $+660$ to $+992$~days, the SEDs are broadly similar in overall shape. SN~2023ixf shows the strongest MIR emission across the full sample, while SN~2024ggi exhibits the smoothest SED, with minimal excess emission from underlying continuum in the 10 and 18~$\mu$m silicate bands. In contrast, SN~2022jox, SN~2021yja, and SN~2023ixf show more pronounced departures from a smooth continuum at these wavelengths, suggestive of developing spectral features. For most objects in this phase range, the SEDs flatten or turn over beyond $\sim$15--20~$\mu$m. The exception is SN~2022jox, which shows a modest upturn at the longest wavelengths. While this could indicate an additional cold dust component, the lack of observations at wavelengths beyond $\sim15$--20~$\mu$m and over longer time baselines limits our ability to determine whether such a component is present \citep{Tinyanont2025_SN2014C, Niculescu-Duvaz2023, Zsiros24}. 

\begin{figure}[tp]
    \centering
    \includegraphics[width=0.45\textwidth]{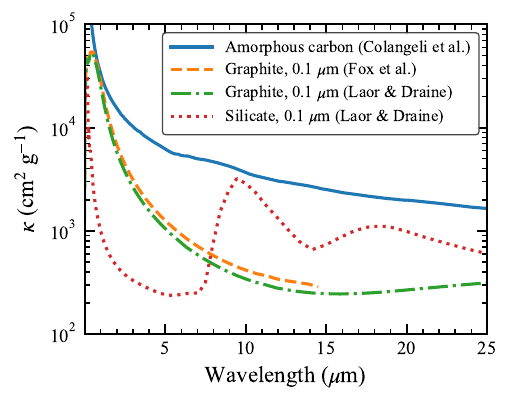}
    \vspace{-6pt}
    \caption{
    Dust opacities, $\kappa$, shown over the JWST/MIRI wavelength range from $5$ to $25~\mu$m. Curves are shown for amorphous carbon from \citet{Colangeli95}, graphite grains with radius $a=0.1~\mu$m from \citet{Fox2011}, and graphite and silicate grains with radius $a=0.1~\mu$m from \citet{Laor93}. 
    }
    \label{fig:kappa_dust}
    \vspace{-12pt}
\end{figure}

Beyond \(\sim\)1000~days, the SEDs slightly diverge. SN~2021gmj, SN~2020jfo, and especially SN~2018cuf show a gradual rise toward the longest wavelengths, whereas SN~2017eaw and SN~2017gmr exhibit excess emission near 10 and 18~\(\mu\)m; SN~2017eaw additionally turns over beyond \(\sim\)20~\(\mu\)m \citep{Shahbandeh23, Pearson2025}, but similar behavior in the other objects is inconclusive because the magnitudes estimated in those bands are largely upper limits. By these epochs, the photospheric contribution is expected to be negligible and the MIR emission is generally interpreted as arising primarily from dust, while radiative transfer models predict that the ejecta become progressively more optically thin at MIR wavelengths as they expand, although the optical depth depends on the dust mass, geometry, and composition \citep{Owen2015, Bevan2016_1987A, Bevan2018, Niculescu-Duvaz2022}. Theoretical models predict that carbon dust condensation commences around these phases in the He/C zones of the ejecta \citep{Sarangi_2013,Sarangi2014, Cherchneff14, Sarangi2022}, potentially contributing to the long-wavelength flux. The exact onset of grain growth is sensitive to the local ejecta conditions, and particularly in SNe with enhanced CSM interaction, dust formation can be accelerated in the dense shell between the forward and reverse shocks \citep{Sarangi2018}, or delayed by reverse-shock heating of the inner ejecta. Unlike the silicate features, amorphous/graphite carbon dust produces a relatively featureless continuum in the IR \citep{Rouleau1991, Laor93, Zubko1996, Colangeli95, Fox2011}, and its presence would manifest primarily as a smooth enhancement of emission at longer wavelengths rather than discrete spectral features, as shown in Figure \ref{fig:kappa_dust}.

In addition to the dust continuum and developing grain features described above, several objects show photometric excesses at wavelengths coincident with the 7.7 and 11.3~$\mu$m PAH bands and the [\ion{Ne}{2}] 12.81~$\mu$m line (shaded regions in Figure~\ref{fig:sed_all}). PAH emission is ubiquitous in the photo-dissociation regions of star-forming environments \citep{Tielens2008, Draine07}, while [\ion{Ne}{2}] emission can originate both in the SN ejecta \citep{Bouchet2006,Milisavljevic2024} where it is predicted to be an important nebular-phase coolant \citep{Roche1993,Dessart2025a} and in surrounding \ion{H}{2} regions within the host galaxy. These excesses are strongest in objects embedded in or projected near luminous star-forming regions, suggesting that host-galaxy contamination remains non-negligible even after PSF subtraction (Appendix~\ref{appendix:host_complexity}). 

\vspace{-1em}
\section{Dust Modeling Methods}\label{sec:modeling}

We model the infrared emission as thermal radiation from dust grains in or near the SN ejecta, which at the epochs probed here ($\sim$400--2300~days) may include newly condensed ejecta dust, dust in a CDS formed between the forward and reverse shocks, and pre-existing circumstellar dust heated by the SN radiation field or ongoing CSM interaction \citep{Barlow2005,Mattila08,Shahbandeh24}. Our modeling procedure is as follows: only filters with S/N $\geq$ 3 are used, and bands that are strongly contaminated by PAH emission (F770W, F1130W) or the 12.8 $\mu$m line (F1280W) are explicitly excluded. NIR photometry, whenever available, is included to anchor the short-wavelength continuum and help constrain the warmer component, which MIRI data alone leave poorly determined. We fit the filter integrated SEDs using an MCMC sampler with analytical modified blackbody prescriptions following the framework of \citet{Hosseinzadeh23}; full details of the model are described in the Appendix of \citet{Pearson2025}. To account for residual systematic uncertainties including unmodeled contributions from forbidden-line emission, we include an intrinsic scatter term $\sigma$ that inflates the photometric uncertainties by a factor of $\sqrt{1 + \sigma^2}$. We use 20 walkers evolved for 2000 steps to ensure convergence, followed by 1000 steps to sample the posterior distributions.

We adopt an optically thin framework for the dust-emitting components in all models. This choice is motivated by the observed SED morphologies: many objects exhibit broad emission around 10 and 18~$\mu$m (Section~\ref{sec:seds}), which is difficult to reproduce with optically thick dust because self-absorption suppresses these features and produces a smoother continuum \citep{Bevan2016_1987A,Niculescu-Duvaz2023}. This assumption is also in line with previous late-time studies of CCSNe that successfully modeled MIR dust emission using optically thin radiative transfer \citep{Shahbandeh23,Zsiros24,Pearson2025, Szalai2013_SN2005af}. We adopt a representative grain radius of $a=0.1~\mu$m for all grain species, as broadband MIR photometry provides only limited constraints on the grain size distribution. This standard choice is supported by previous studies showing only a weak dependence of the inferred dust properties on grain size \citep{Fox2011,Shahbandeh23,Pearson2025}. A systematic exploration of any grain size effects is left to future work. We consider three grain species motivated by dust condensation models for CCSNe: astronomical silicates ($\rho_{\rm sil}=3.3$~g~cm$^{-3}$; \citealt{Laor93}), amorphous carbon ($\rho_{\rm aC}=2.5$~g~cm$^{-3}$; \citealt{Colangeli95}), and graphite ($\rho_{\rm gr}=2.26$~g~cm$^{-3}$; \citealt{Draine07}). For each SN, we compare a set of physically motivated model configurations spanning single and multi component prescriptions and different dust compositions. Our preferred model is selected based on goodness of fit and maximum likelihood, while favoring the simplest model when additional components do not produce a statistically meaningful improvement.

At $\sim$400--900~days post-explosion (SN~2022wsp, SN~2022acko, SN~2024ggi, SN~2022jox, and SN~2021yja), contemporaneous NIR photometry is generally unavailable, leaving the warm/hot emission component poorly constrained. We therefore represent this emission with a simple blackbody which we fit, that serves as an empirical representation of the residual short wavelength flux rather than a physically distinct dust component. At these epochs the ejecta are still evolving through the transition from optically thick to optically thin, so the blackbody may capture a combination of fading SN emission and/or very warm material at small radii; the optically thin treatment is applied only to the dust components at longer wavelengths. We adopt a BB$+$Si model (hereafter Si\_BB), in which the blackbody accounts for the poorly constrained hot continuum and the optically thin silicate component reproduces the emerging dust components. Silicate grains are a physically motivated choice for the youngest objects, as they are expected to form early in O-rich ejecta \citep{Sarangi_2013, Cherchneff14,Sarangi_Chercheneff2015}.

By $\sim$900--1900~days, the photospheric contribution has largely faded, and the SED in SN~2021gmj, SN~2020jfo, and SN~2018cuf is better reproduced by a C$+$Si model (hereafter CSi). However, the hot/warm component is constrained by only a single photometric point in several objects at this phase, precluding robust constraints on its temperature or composition. We therefore interpret the CSi fits only as evidence for an additional warm component, without assigning a specific physical origin.


SN~2023ixf at $+$992~days requires a more complex configuration. Its exceptional MIR luminosity, NIR photometry, and evidence for dense and structured CSM from the progenitor \citep{Kilpatrick_2023ixf,WJG_2025_SN2023ixf, Bostroem2024_SN2023ixf,Bostroem2025_latetime_HST_23ixf} together motivate a three-component 2C$+$Si model (hereafter 2CSi): two carbon components at distinct temperatures plus a silicate component. A similar model prescription was recently adopted for SN~2023ixf by \citet{Singh2026_23ixf}. 

\begin{figure*}[ht]
	\centering
\includegraphics[width = 0.95\textwidth]{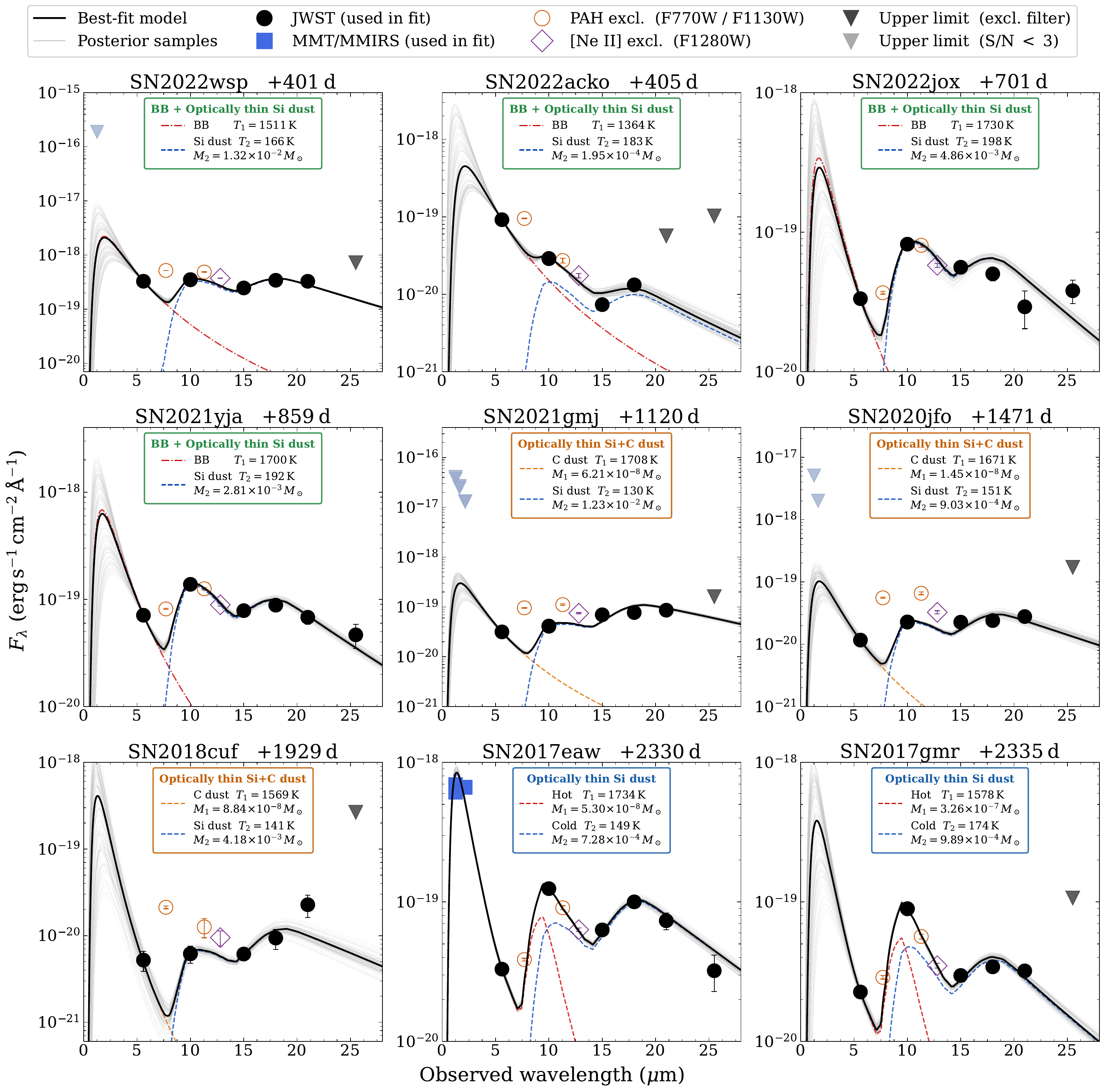}
    \caption{JWST/MIRI spectral energy distributions and best-fit dust models for nine SNe in the sample, arranged in order of increasing phase from left to right and top to bottom. Filled circles show the observed MIRI photometry with $1\sigma$ error bars; open symbols indicate filters excluded from the fit due to suspected PAH (F770W, F1130W) or [\ion{Ne}{2}] (F1280W) contamination. Solid curves show the total best-fit model, with individual components (blackbody, hot dust, cool silicate) plotted separately.}
    \label{fig:sed_model_all}
\end{figure*}

At later epochs ($\gtrsim$2300~days; SN~2017eaw and SN~2017gmr), the MIR-emitting ejecta are expected to be more optically thin, and the silicate features become clearly visible \citep{Shahbandeh23,Pearson2025}. These older SEDs are well reproduced by a two-component silicate model (hereafter 2Si). The best-fit parameters for the full sample are shown in Figures~\ref{fig:sed_model_all} and~\ref{fig:sed_23ixf_24ggi} and summarized in Table~\ref{tab:fit_params}.

\vspace{-1.4em}

\begin{figure*}[tp]
\centering 
\includegraphics[ width=0.76\textwidth]{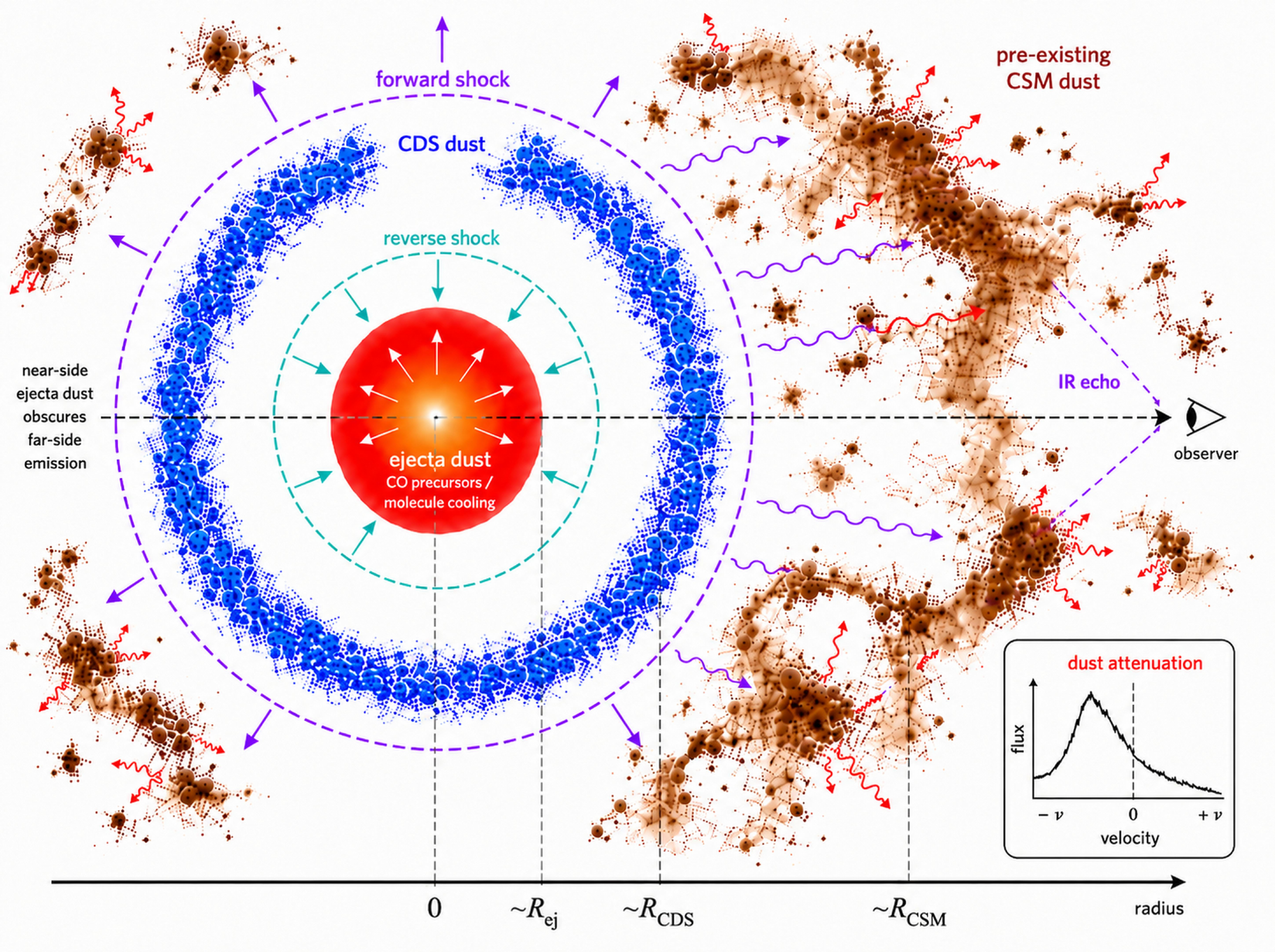} 
\caption{Schematic of the dust origin channels in CCSNe. Dust may form in the expanding ejecta, aided by molecule formation and cooling through species such as CO, or in a CDS produced by ejecta-CSM interaction. MIR emission can also arise from pre-existing CSM dust formed by progenitor mass loss and heated by the SN flash and/or ongoing interaction, producing an IR echo. The radii $\sim R_{\rm ej}$, $\sim R_{\rm CDS}$, and $\sim R_{\rm CSM}$ mark the characteristic ejecta, CDS/shock, and pre-existing CSM dust scales. The inset shows the expected line-profile asymmetry when dust preferentially attenuates far-side ejecta emission. } \label{fig:dust_origins_schematic} 

\end{figure*}

\section{Results and Discussion}\label{sec:results_discussions}

\begin{figure*}[htp]
	\centering
\includegraphics[width = 0.8\textwidth]{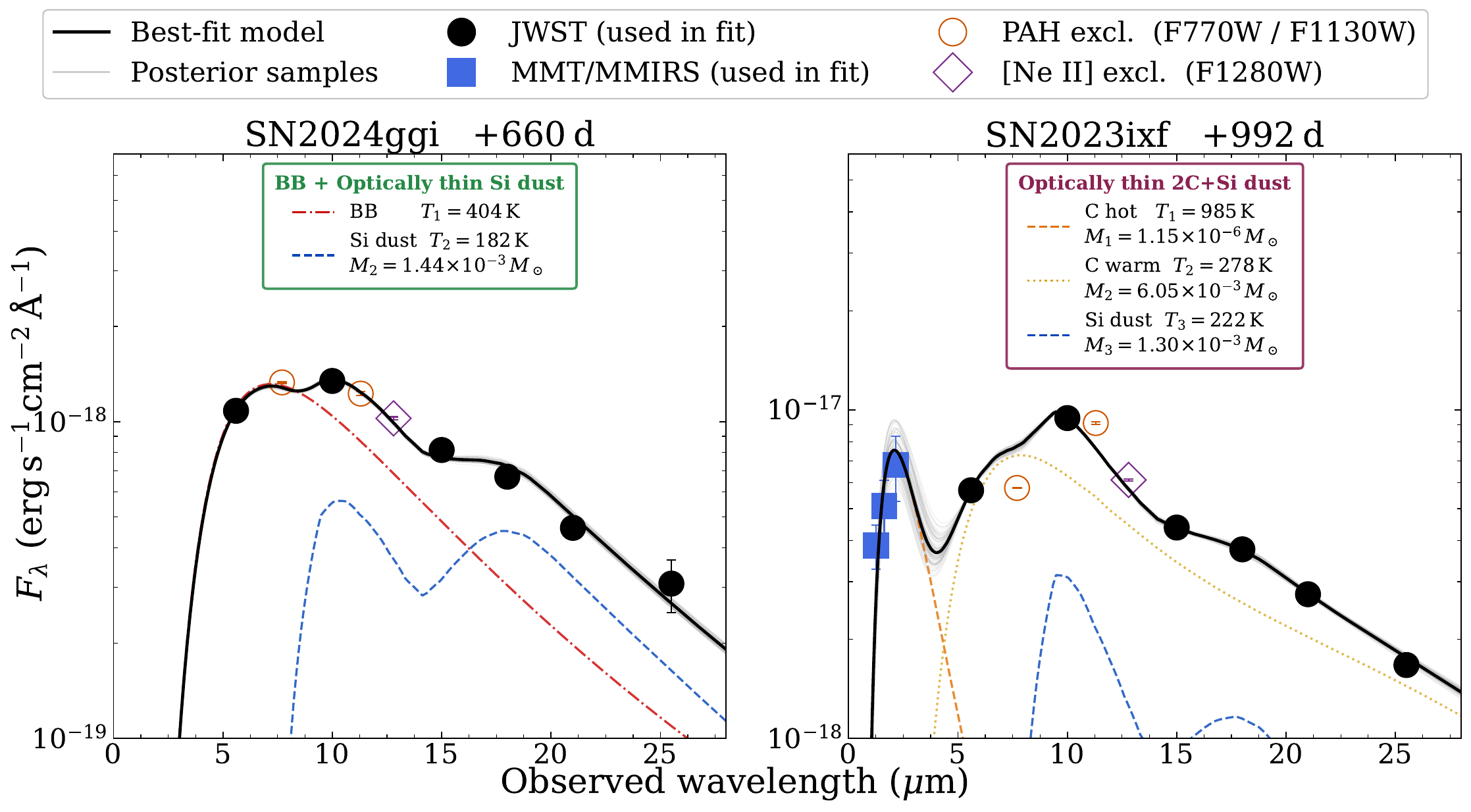}
    \caption{JWST/MIRI best fit dust models for SN~2024ggi ($+$660~d) and SN~2023ixf ($+$992~d).}
    \label{fig:sed_23ixf_24ggi}\label{2x2_grid}
\end{figure*}

\begin{figure*}[htp]
 	\centering
 \includegraphics[width = 0.9\textwidth]{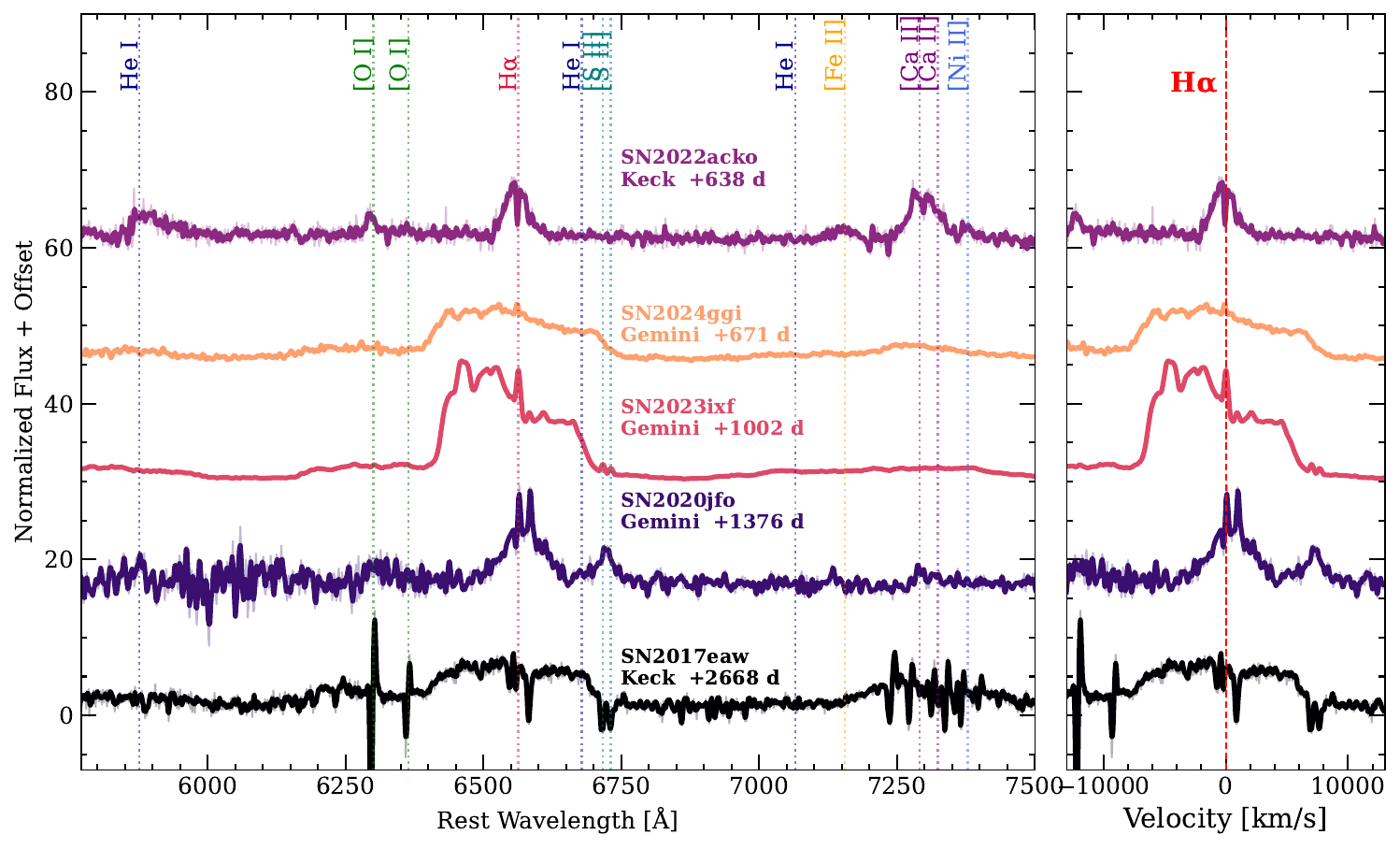}
    \caption{Late-time optical spectra of five SNe in the sample obtained with Keck/LRIS and Gemini/GMOS, ordered from bottom to top by increasing phase. \textit{Left:} Normalized spectra in the 5700--7500~\AA\ range, with key emission lines labeled. \textit{Right:} H$\alpha$ profiles plotted in velocity space. SN~2023ixf and SN~2024ggi display blue-red asymmetries indicative of internal dust in the ejecta, while SN~2017eaw shows a boxy, flat-topped profile characteristic of dust likely arising in the CDS.}
    \label{fig:all_spectra_detection}
 \end{figure*}

In the following subsections, we present the dust properties of each SN followed by implications for dust origins and progenitor scenarios, CO detections, late-time spectroscopic diagnostics, sample-wide trends, the influence of CSM and any correlations with SN explosion properties. Throughout this section, dust temperatures, dust masses, and fluxes refer to the best-fit values from the SED modeling (Fig.~\ref{fig:sed_model_all}; Fig.~\ref{fig:sed_23ixf_24ggi}; Table~\ref{tab:fit_params}), with the corresponding photometry in Table~\ref{tab:miri_photometry}. We show the late-time optical spectra of five SNe in the sample with Keck/LRIS and Gemini/GMOS (Figure ~\ref{fig:all_spectra_detection}). 

The MIR excess in CCSNe at $\sim$1--7~yr can arise from predominantly three channels: (i)~newly condensed dust in the freely expanding, metal-rich ejecta \citep{Kozasa1989, Todini01, Nozawa2003, Sarangi2014, Sarangi18}; (ii)~newly condensed dust in the CDS that forms at the ejecta-CSM interface when the SN drives a radiative shock into dense CSM \citep{Pozzo04, Smith2008_SN2006jc, Sarangi2018, Sarangi2022, Takei2025}; and (iii)~pre-existing circumstellar dust formed in the progenitor wind and radiatively heated by the SN flash or the interaction luminosity, which at large radii manifests as an IR echo \citep{Dwek1987, Bode80, Fox2010_05ip, Fox2011, Hosseinzadeh2022}. The dust origin scenarios are summarized in  Figure \ref{fig:dust_origins_schematic}. 

Distinguishing these dust origin channels is challenging from a single SED; instead, the literature combines several physically motivated consistency checks, dust temperature and composition, inferred or equilibrium radii compared to the ejecta and shock radii, CO detections as a chemical precursor to grain condensation, the temporal evolution of dust mass and temperature, and optical line-profile diagnostics such as blue-red asymmetries, intermediate-width components, and boxy profiles \citep{Lucy89, Bevan2016_1987A, Dessart2025_dusty_typeII}. 

We adopt the same multi-faceted approach here. A widely used geometric check compares the dust location to the characteristic radii of the system. For optically thin grains in radiative equilibrium with a central heating source of luminosity $L_{\rm heat}$, the equilibrium radius is estimated by \(R_{\rm eq} = \left(L_{\rm heat}/16\pi\sigma T_{\rm d}^{4}\right)^{1/2}\), 
where $T_{\rm d}$ is the fitted dust temperature. This is an external diagnostic, not an output of our optically thin SED modeling, and its dominant uncertainty is $L_{\rm heat}$, which evolves with phase: early epochs are powered by the SN luminosity, while nebular heating is primarily powered by radioactive deposition, and CSM interaction possibly becoming a dominant heating mechanism at much later times in normal Type II SNe. We now present the 11 objects in order of increasing phase to highlight the evolutionary progression of their MIR emission.

\indent\textbf{SN~2022wsp (+401~d):} \citet{SN2022wsp_Vasylev} found no flash-ionization features in HST ultraviolet spectra obtained $\sim10$~days after explosion, but derived a relatively high total reddening of $E(B-V)\approx0.35$~mag. However, the earliest available classification spectrum may display a ledge-like H$\alpha$ profile \citep{TNS_22wsp}. At +401~d, the MIRI SED rises from F560W $\approx 34~\mu$Jy to F2100W $\approx 483~\mu$Jy. A hot blackbody ($T_{BB} \sim 1511$~K) plus cool silicate ($T_{cool} \sim 166$~K) fit returns
$M_{\rm cool} \sim 1.32\times10^{-2}~M_{\odot}$. 

No constrained $M(^{56}\mathrm{Ni})$ is available for this object, so we cannot anchor the heating to radioactive decay and instead adopt a generic $L_{\rm heat}\sim10^{38}\text{--}10^{39}\,\mathrm{erg\,s^{-1}}$. With $T_{\rm cool} \sim  166\,\mathrm{K}$ this gives $R_{\rm eq}\sim(7\text{--}22)\times10^{15}\,\mathrm{cm}$. No late-time spectrum is available, so we lack a measured nebular ejecta velocity; the early photospheric velocity of $\approx7500\text{--}8500\,\mathrm{km\,s^{-1}}$ \citep{SN2022wsp_Vasylev} provides an upper bound on the nebular value, corresponding to $R_{\rm ej}\lesssim(26\text{--}29)\times10^{15}\,\mathrm{cm}$, suggesting newly formed dust in the ejecta.

An IR echo, however, would be governed by the earlier SN radiation field rather than the contemporaneous radioactive luminosity. At $+401$\,d, the characteristic light-travel scale is $ct/2\simeq5.2\times10^{17}\,\mathrm{cm}$, and dust heated by a peak luminosity of order $10^{42}$--$10^{43}\,\mathrm{erg\,s^{-1}}$ could reach equilibrium temperatures comparable to the fitted cold component at radii much larger than the ejecta. In the absence of late-time spectroscopy and CO constraints, we do not assign a unique dust origin for SN~2022wsp; ejecta, circumstellar, and echo contributions remain possible and may coexist.

\textbf{SN~2022acko ($+$405~d).}
\citet{22acko_Bostroem} obtained the earliest far-UV spectra of any SN~IIP to date, finding a low-luminosity, low-velocity explosion with no flash features. \citet{VanDyk2023} identified a candidate RSG progenitor consistent with $M_{\rm ZAMS}\lesssim 9~M_{\odot}$. The first JWST NIRSpec and MIRI spectra of any core-collapse SN, obtained at $+$50~d by \citet{Shahbandeh2024_22acko}, detected neither CO nor dust and placed an upper limit on the CO mass of $<10^{-8}~M_{\odot}$. Subsequent JWST NIRSpec/MIRI spectroscopy at $+$259 and $+$368~d detected CO with a mass increasing from $\sim1.6\times10^{-4}$ to $2.5\times10^{-4}~M_{\odot}$, but still found no evidence for SiO or dust emission, suggesting delayed or inefficient dust formation in this low-mass explosion \citep{Medler2026}. Nebular-phase analyses by \citet{Teixeira2026} and \citet{Lin_22acko_2025} converge on $M_{\rm ZAMS} \approx 9$--$10~M_{\odot}$ and $M(^{56}\mathrm{Ni})=0.014\pm 0.004~M_{\odot}$. At $+$405~d, the MIRI SED is flat and faint ($\sim$5--19~$\mu$Jy across the imaging filters). Our fit returns a blackbody at $T_{\rm BB} \sim 1364$~K and an optically thin silicate component at $T_{\rm cool} \sim 183$~K with $M_{\rm cool} \sim 1.95 \times10^{-4}~M_{\odot}$. This mass is consistent with chemical kinetic predictions for in-situ ejecta dust at $\sim$1~yr post-explosion \citep{Sarangi18}, and defines the lower envelope of the sample at early phases.

From $M(^{56}\mathrm{Ni})=0.014\,M_{\odot}$ \citep{Teixeira2026, Lin_22acko_2025}, the $^{56}$Co heating rate at $+$405~d is $L_{\rm Co}\approx(0.5$--$5)\times10^{39}~{\rm erg~s^{-1}}$ (positron-escape to full trapping). With $T_{\rm cool} \sim 183$~K, the central-source equilibrium radius is $R_{\rm eq}\approx(13$--$40)\times10^{15}$~cm, comparable to the ejecta radius $R_{\rm ej}\approx9\times10^{15}$~cm from the H$\alpha$ half-width ($\approx2500~{\rm km~s^{-1}}$; Fig.~\ref{fig:all_spectra_detection}). Because heating is distributed ($^{56}$Co mixed in ejecta) rather than central, $R_{\rm eq}$ serves only as a consistency check: agreement with $R_{\rm ej}$ shows the dust temperature is energetically plausible for grains in the ejecta. The low dust mass ($M_{\rm d}=1.95\times10^{-4}\,M_{\odot}$) and compact warm component ($T_{\rm BB}=1360$~K, $R_{\rm BB}\approx5\times10^{14}$~cm) are broadly consistent with emission from within or near the ejecta, though not uniquely diagnostic. The nearly symmetric H$\alpha$ profile (see Figure~\ref{fig:all_spectra_detection}) implies weak differential attenuation. For a simple foreground-screen geometry, external dust would attenuate the red and blue wings similarly, so a marked asymmetry would more naturally indicate dust mixed with, or internal to, the emitting region. The near symmetry therefore favors low optical depth, clumpy geometry, or emission from less obscured regions, although an internal ejecta contribution remains possible and a circumstellar component cannot be excluded.

\textbf{SN~2024ggi ($+$660~d).}
SN~2024ggi was discovered by ATLAS \citep{Tonry2024} on 2024 April 11 (MJD~60411.14). Flash-ionization features of H~I, He~I, He~II, C~III, C~IV, and N~III--V emerged within the first day and persisted for $3.8\pm1.6$~d \citep{Shrestha2024, WJG_2024, Pessi2024, Zhang2024}. Modeling of the early spectra and bolometric light curve implies a RSG progenitor surrounded by a dense, confined CSM with $\dot{M} \sim 10^{-3}$--$10^{-2}~M_{\odot}\,\mathrm{yr}^{-1}$ (assuming a $10$~km~s$^{-1}$ wind) extending to $\lesssim 5\times 10^{14}$~cm \citep{Shrestha2024, WJG_2024, Zhang2024, Ertini2025, Amar2025_SN2024ggi}. Pre-explosion \textit{HST} and \textit{Spitzer} imaging identified a $\sim13~M_{\odot}$ red supergiant progenitor with $R\approx890~R_{\odot}$ \citep{Xiang2024}. Hydrodynamical modeling of the bolometric light curve favors $M_{\rm ZAMS}=15~M_{\odot}$, a pre-SN radius of $517~R_{\odot}$, and an explosion energy of $\sim 1.2\times 10^{51}$~erg, with $M(^{56}\mathrm{Ni})\lesssim 0.035~M_{\odot}$ \citep{Ertini2025}; nebular analyses subsequently revise the nickel mass upward to $\sim 0.05$--$0.06~M_{\odot}$ \citep{Ferrari25_SN2024ggi, Dessart2025_SN2024ggi}. Nebular-phase optical and NIR spectroscopy independently estimate $M_{\rm ZAMS} \approx 14~M_{\odot}$ \citep{Hueichapan2026_SN2024ggi}. The first JWST epoch, with NIRSpec and MIRI/LRS observations, at $+$55~d revealed neither dust nor CO emission, with the 0.4--21~$\mu$m SED matching a PHOENIX/1D photospheric model \citep{Baron2025}. CO first overtone emission was subsequently detected at $+$250--319~d \citep{Hueichapan2026_SN2024ggi} and again at $+$285 and $+$385~d \citep{Mera2026_SN2024ggi}. By $+$660~d, the MIRI SED is bright and rises smoothly from F560W $\approx 113~\mu$Jy to a peak of F1800W $\approx 725~\mu$Jy, with F2100W $\approx 680~\mu$Jy. Late time nebular spectra up to $\sim$ 400 days show no ongoing narrow line CSM interaction \citep{Ferrari25_SN2024ggi, Hueichapan2026_SN2024ggi}.

The best-fit dust model includes a warm blackbody component ($T_{\rm BB} \sim 404\,$K, $R_{\rm BB}\approx 9.9\times10^{4}~R_{\odot}$) and an optically thin silicate component ($T_{\rm cool} \sim 182 $~K, $M_{\rm cool} \sim 1.44 \times10^{-3}~M_{\odot}$), as shown in Figure \ref{fig:sed_23ixf_24ggi}. The warm blackbody temperature and large radius are consistent with emission from heated pre-existing circumstellar dust at the IR echo radius, rather than residual photospheric emission. Adopting $M(^{56}\mathrm{Ni})=0.055\,M_{\odot}$ \citep{Ferrari25_SN2024ggi, Dessart2025_SN2024ggi}, the $^{56}$Co heating rate at +660\,d is $L_{\rm Co}\sim(0.2\text{--}2)\times10^{39}\,\mathrm{erg\,s^{-1}}$, which for $T_{\rm d}=182\,\mathrm{K}$ implies $R_{\rm eq}\sim(8\text{--}25)\times10^{15}\,\mathrm{cm}$. This is broadly comparable to $R_{\rm ej}\approx26\times10^{15}\,\mathrm{cm}$ inferred from the H$\alpha$ half-width ($\approx4500\,\mathrm{km\,s^{-1}}$; Fig.~\ref{fig:all_spectra_detection}). The cooler silicate component ($T_{\rm cool} \sim 182$~K), with $L_{\rm Co}\approx2\times10^{39}~{\rm erg~s^{-1}}$ at $+$660~d giving $R_{\rm eq}\approx(18$--$31)\times10^{15}$~cm near $R_{\rm ej}$, plus the CO detection at $+$285--385~d \citep{Mera2026_SN2024ggi} and a subtle blue-red H$\alpha$ asymmetry, points to newly forming dust interior to the line-emitting region; SN~2024ggi is therefore best read as a composite of a pre-existing echo and a growing ejecta/CDS component.

\textbf{SN~2022jox ($+$701~d).}
\citet{22jox_Andrews2024} reported high-cadence optical/UV observations beginning $\sim$0.75~d after explosion. Strong but short-lived flash features (\ion{H}{1}, \ion{He}{2}, \ion{C}{4}, \ion{N}{4}) imply $\dot{M} \sim 10^{-3}$--$10^{-2}~M_{\odot}\,\mathrm{yr}^{-1}$. A multi-component H$\alpha$ profile at $\sim$200~d indicates an outer asymmetric CSM in addition to the inner shell. SN~2022jox peaked at $M_V \approx -17.3$ and synthesized $M(^{56}\mathrm{Ni}) \approx 0.04~M_{\odot}$, otherwise typical for SNe~IIP. At $+$701~d, the MIRI SED rises from F560W $\approx 3.5~\mu$Jy to F2550W $\approx 83~\mu$Jy. A blackbody ($T_{\rm BB} \sim 1730$~K) plus optically thin silicate ($T_{\rm cool} \sim 198$~K) fit yields a dust mass of $M_{\rm cool} \sim 4.86 \times10^{-3}~M_{\odot}$.

With $M(^{56}\mathrm{Ni})=0.04\,M_{\odot}$ \citep{22jox_Andrews2024}, $^{56}$Co heating gives $L_{\rm Co}\approx(0.3\text{--}1)\times10^{39}\,\mathrm{erg\,s^{-1}}$, so with $T_{\rm cool} \sim 198\,\mathrm{K}$, $R_{\rm eq}\approx(8\text{--}15)\times10^{15}\,\mathrm{cm}$. The early H$\alpha$ envelope velocity ($\approx7000\,\mathrm{km\,s^{-1}}$; \citealt{22jox_Andrews2024}) gives $R_{\rm ej}\approx42\times10^{15}\,\mathrm{cm}$, so $R_{\rm eq}$ lies interior to it, given the assumed luminosity. If the assumed heating luminosity is higher than the $^{56}$Co rate, or if there is additional luminosity from CSM interaction or radiative heating of pre-existing dust, the inferred $R_{\rm eq}$ would increase and move outward, in which case a pre-existing or circumstellar contribution becomes more accountable. A CDS contribution alongside ejecta condensation is also possible, though without contemporaneous CO or late-time spectroscopy this cannot be confirmed.

\textbf{SN~2021yja ($+$859~d).}
\citet{Hosseinzadeh2022} placed a stringent upper limit of $M_{\rm ZAMS}\lesssim 9~M_{\odot}$ from archival HST imaging and inferred a relatively low mass loss rate of $\dot{M} \sim 10^{-6}~M_{\odot}\,\mathrm{yr}^{-1}$. However, the properties and extent of the CSM surrounding SN~2021yja remain poorly constrained and are still the subject of debate. \citet{Kozyreva2022_SN2021yja} required low density CSM to fit the rapid rise time and the X-ray/radio detections, while \citet{Vasylyev2024_SN2021yja} and \citet{Nagao2024_SN2021yja} reported unusually high continuum polarization, indicating an aspherical extended H envelope. At $+$859~d, the MIRI SED rises from F560W $\approx 7.4~\mu$Jy to F2550W $\approx 101~\mu$Jy. A blackbody ($T_{\rm BB} \sim 1700$~K) plus silicate fit yields $T_{\rm cool} \sim 192$~K and $M_{\rm dust} \sim 2.81 \times10^{-3}~M_{\odot}$. 

From literature estimates, $M(^{56}\mathrm{Ni})\approx0.13~M_{\odot}$, $^{56}$Co heating gives $L_{\rm Co}\approx(0.3$--$1)\times10^{39}~{\rm erg~s^{-1}}$, so with $T_{\rm cool} \sim 192$~K, $R_{\rm eq}\approx(9$--$16)\times10^{15}$~cm. We have no late-time spectrum for this object; adopting a generic nebular ejecta velocity of $\approx3000$--$4000~{\rm km~s^{-1}}$ gives $R_{\rm ej}\approx(22$--$30)\times10^{15}$~cm, so $R_{\rm eq}$ is comparable to $R_{\rm ej}$. With both the luminosity and the velocity assumed rather than measured, this is only a weak constraint with large uncertainties. SN~2021yja shows high continuum polarization suggesting departures from spherical symmetry  \citep{Vasylyev2024_SN2021yja, Nagao2024_SN2021yja}, and therefore the dust masses inferred from spherical models ($2.81\times10^{-3}~M_{\odot}$) may represent lower limits. We therefore do not assign a dust origin for SN~2021yja, leaving all three channels open.

\textbf{SN~2023ixf ($+$992~d).}
Pre-explosion HST and Spitzer imaging of the M101 site reveal a dusty, variable red supergiant progenitor \citep{VanDyk2024_SN2023ixf, Kilpatrick_2023ixf, Li2024_SN2023ixf_dusty, Jencson2023_SN2023ixf, Soraisam2023_SN2023ixf}, with $M_{\rm ZAMS}$ estimates spanning $\sim$9--22~$M_{\odot}$ \citep{Ransome_2023ixf, Hsu2025_SN2023ixf, Moriya2024_SN2023ixf, Vink2025_SN2023ixf, Bersten2024_SN2023ixf, Ragosta2026_23ixf_dust, Martinez2024_SN2023ixf, Dessart2026}. Flash spectroscopy within the first day requires a dense, confined CSM \citep{Bostroem23_23ixf,Bostroem2024_SN2023ixf, Smith2023_2023ixf, WJG_2023_2023ixf, Chandra2024_SN2023ixf, Zimmeman_SN2023ixf, Shrestha2025_SN2023ixf, Teja2023_SN2023ixf, Bostroem2025_latetime_HST_23ixf}. NIR spectroscopy detected CO first-overtone emission between $+199$ and $+307$~d, with CO velocities of $\sim3000$--$3500~{\rm km~s^{-1}}$ and warm dust masses of order $\sim10^{-5}~M_{\odot}$, supporting an emerging internal dust component in the ejecta \citep{Park2025_SN2023ixf}. Plateau-phase JWST spectra at $+$33.6~d show no CO or dust \citep{Derkacy2026_23ixf}; the panchromatic JWST monitoring program from $+$250~d to $+$720~d \citep{Medler2025_SN2023ixf} subsequently detected the CO first-overtone and fundamental bands and revealed an evolving IR excess. \citet{Singh2026_23ixf} combined multi-wavelength data over $+$150--750~d to disentangle the early circumstellar IR echo from newly forming internal dust, finding a growing silicate component reaching $\sim 2\times10^{-3}~M_{\odot}$ by $+$750~d. At $+$992~d, our MIRI SED is the brightest in the sample: F560W $\approx 595~\mu$Jy rises to F1130W $\approx 3881~\mu$Jy, with a secondary plateau at F1800W--F2100W $\approx 4040$--$4065~\mu$Jy. 

The JWST/MIRI SED requires a three-component 2CSi model as seen in Figure \ref{fig:sed_23ixf_24ggi}: hot graphite ($T_{\rm C,hot} \sim 985$~K, $M_{\rm C,hot} \sim 1.15 \times10^{-6}~M_{\odot}$), cool graphite ($T_{\rm C,cool} \sim 278$~K, $M_{\rm C,cool} \sim 6.05 \times10^{-3}~M_{\odot}$), and silicate ($T_{\rm Si, cool} \sim 222$~K, $M_{\rm Si, cool}=1.30 \times10^{-3}~M_{\odot}$). The total cool-dust mass (graphite $+$ silicate) is $\approx 7.4\times10^{-3}~M_{\odot}$, consistent with continued growth since the $+$750~d epoch  \citep{Singh2026_23ixf}. The mixed carbon-silicate composition and the three component model distinguish SN~2023ixf from all other SNe in the sample and may reflect the higher mass of condensable material in the ejecta and the favorable conditions created by the dense CSM interaction.

SN~2023ixf ($T_{\rm Si} \sim 222$~K) is past the point where $^{56}$Co can heat the dust ($L_{\rm Co}\approx10^{38}~{\rm erg~s^{-1}}$ at $+$992~d), and its heating is instead set by ongoing interaction, measured at $L_{\rm H\alpha}\approx4.2\times10^{38}~{\rm erg~s^{-1}}$ at $+$445~d; adopting $L_{\rm heat}\approx(4$--$10)\times10^{38}~{\rm erg~s^{-1}}$ gives $R_{\rm eq}\approx(8$--$12)\times10^{15}$~cm. Our Gemini spectrum at $+$1002~d (Fig.~\ref{fig:all_spectra_detection}) resolves central H$\alpha$ peaks at $\approx1500~{\rm km~s^{-1}}$ from the radioactively powered ejecta and boxy edges at $\approx8000~{\rm km~s^{-1}}$ from the CSM interaction, giving $R_{\rm ej,inner}\approx13\times10^{15}$~cm and $R_{\rm ej,outer}\approx69\times10^{15}$~cm; $R_{\rm eq}$ thus coincides with the inner ejecta and lies well inside the interaction shell. The CO detection at $+$250--720~d \citep{Medler2025_SN2023ixf}, the pronounced blue-red H$\alpha$ asymmetry indicating dust internal to the line-emitting region \citep{Dessart2025_dusty_typeII}, the mixed graphite-silicate composition expected from condensation in chemically stratified post-shock gas \citep{Sarangi2022}, and a mass increase from $\sim2\times10^{-3}~M_{\odot}$ at $+$750~d to $\sim7\times10^{-3}~M_{\odot}$ at $+$992~d (a comparison across different compositions, but in the sense expected for active condensation) together make newly formed CDS dust the favored origin, with the early circumstellar echo a distinct, earlier-phase component \citep{Singh2026_23ixf}.

\textbf{SN~2021gmj ($+$1120~d).}
An intermediate-luminosity SN~IIP with $M_V \approx -15.5$. \citet{Murai2024_SN2021gmj} and \citet{21gmj_Meza_Retamal_2024} converge on $M_{\rm ZAMS}\approx 12~M_{\odot}$ from nebular [\ion{O}{1}]/[\ion{Ca}{2}], $M(^{56}\mathrm{Ni}) \approx 0.014$--$0.02~M_{\odot}$, and an explosion energy $\approx 3\times 10^{50}$~erg. A fast early light-curve rise and a broad ledge feature near 4600~\AA\ in the first spectra  indicate confined CSM with a mass of $\sim 0.025~M_{\odot}$ \citep{21gmj_Meza_Retamal_2024}. At $+$1120~d, the MIRI SED rises to F2100W $\approx 127~\mu$Jy. An amorphous-carbon plus silicate fit returns a hot carbon component at $T_{\rm hot} \sim 1710$~K with $M_{\rm hot} \sim 6.21 \times10^{-8}~M_{\odot}$, and a cool silicate component at $T_{\rm cool} \sim 130$~K with $M_{\rm cool}\sim 1.23 \times10^{-2}~M_{\odot}$. 
The pre-explosion mass-loss rate for the progenitor of SN~2021gmj is estimated to be $\gtrsim 10^{-3}~M_{\odot}~\mathrm{yr}^{-1}$, with some models placing it up to roughly $\sim 10^{-2}~M_{\odot}~\mathrm{yr}^{-1}$ in the final years immediately before core collapse \citep{Murai2024_SN2021gmj, 21gmj_Meza_Retamal_2024}. 

Adopting a generic late-time heating luminosity of $L_{\rm heat}\sim(1\text{--}10)\times10^{38}\,\mathrm{erg\,s^{-1}}$ and the fitted temperature $T_{\rm cool} \sim 130\,\mathrm{K}$ gives $R_{\rm eq}\approx(11\text{--}35)\times10^{15}\,\mathrm{cm}$, comparable to the ejecta scale $R_{\rm ej}\approx24\times10^{15}\,\mathrm{cm}$ inferred for an assumed ejecta velocity of $\approx2500\,\mathrm{km\,s^{-1}}$ \citep{Murai2024_SN2021gmj}. For comparison, if the observed MIR emission at +1120\,d arises from an IR echo, the dust must lie at $R_{\rm echo} \gtrsim ct/2 \approx 1.5\times10^{18}\,\mathrm{cm}$, corresponding to a much more distant pre-existing dust shell. The confined CSM inferred for this object \citep{21gmj_Meza_Retamal_2024} makes a CDS contribution possible, however, the scenario for an IR echo cannot be completely ruled out.

\textbf{SN~2020jfo ($+$1471~d).}
\citet{Sollerman2021_SN2020jfo} showed that SN~2020jfo exhibited an unusually short ($\sim$60--65~d) plateau and an ejecta mass of $\sim$5~$M_{\odot}$. The inferred $^{56}$Ni mass is $M(^{56}\mathrm{Ni})\approx0.025$--$0.033~M_{\odot}$, with a nebular-phase progenitor mass of $\sim$12~$M_{\odot}$ \citep{Teja2022_SN2020jfo,Kilpatrick2023_SN2020jfo}. There is no robust evidence for dense CSM interaction. At $+$1471~d, the MIRI SED is the faintest among the older objects: F560W $\approx 1.2~\mu$Jy rises to F2100W $\approx 41~\mu$Jy. A carbon plus silicate fit yields a hot carbon component at $T_{\rm hot} \sim 1670$~K ($M_{\rm hot} \sim 1.45 \times10^{-8}~M_{\odot}$) and a cool silicate component at $T_{\rm cool} \sim 152$~K with $M_{\rm cool} \sim 9.03\times10^{-4}~M_{\odot}$, an order of magnitude below SN~2021gmj at a comparable phase.

From the nebular spectrum we measure an H$\alpha$ half-width of $\approx3000\,\mathrm{km\,s^{-1}}$ (Fig.~\ref{fig:all_spectra_detection}), corresponding to an ejecta scale of $R_{\rm ej}\approx vt\approx38\times10^{15}\,\mathrm{cm}$. The fitted dust temperature is $T_{\rm cool}\sim 152\,\mathrm{K}$; for an adopted interaction luminosity of $L_{\rm heat}\sim(1\text{--}5)\times10^{38}\,\mathrm{erg\,s^{-1}}$, this implies $R_{\rm eq}\approx(8\text{--}18)\times10^{15}\,\mathrm{cm}$, somewhat interior to $R_{\rm ej}$. The narrow, symmetric H$\alpha$ profile suggests limited dust-induced attenuation along the line of sight. Taken together, the data are consistent with dust associated with the ejecta or a dense shell, although pre-existing dust from CSM cannot be excluded on the basis of the available diagnostics alone, given there can be large uncertainties in the assumed interaction strength and luminosities.

\textbf{SN~2018cuf ($+$1929~d).}
\citet{2018cuf_Yize} derived $M_{V,\rm max}=-16.73\pm 0.32$, $M(^{56}\mathrm{Ni})=0.04\pm 0.01~M_{\odot}$, $M_{\rm ZAMS}\approx 14.5~M_{\odot}$, and required $\sim 0.07~M_{\odot}$ of confined CSM to fit the early light curve. High-velocity H$\alpha$ features at $\sim$11{,}000~km~s$^{-1}$ during the plateau support sustained ejecta--CSM interaction. At $+$1929~d, the MIRI SED is faint at short wavelengths (F560W $\approx 0.55~\mu$Jy) but rises sharply to F2100W $\approx 33~\mu$Jy, leaving an excess that is not fully captured by the simplest dust prescriptions. A graphite plus silicate fit yields a hot graphite component at $T_{\rm hot} \sim 1570$~K ($M_{\rm hot}=8.84\times10^{-8}~M_{\odot}$) and a cool silicate component at $T_{\rm cool} \sim 141$~K with $M_{\rm cool} = 4.18\times10^{-3}~M_{\odot}$.

With $^{56}$Co contribution negligible at this epoch, we adopt $L_{\rm heat}\sim(3\text{--}10)\times10^{38}\,\mathrm{erg\,s^{-1}}$; for $T_{\rm cool} \sim 141\,\mathrm{K}$ this gives $R_{\rm eq}\approx(16\text{--}30)\times10^{15}\,\mathrm{cm}$. We have no late-time spectrum; an assumed velocity of $\approx3000\,\mathrm{km\,s^{-1}}$ gives $R_{\rm ej}\approx5.0\times10^{16}\,\mathrm{cm}$, leaving $R_{\rm eq}$ interior. The reported high-velocity H$\alpha$ may be consistent with sustained interaction \citep{2018cuf_Yize} and makes a CDS contribution plausible alongside any ejecta condensation. A pre-existing dust echo is also possible, with a detectable echo persisting to $+1929$\,d would require dust at a radius of order $R_{\rm echo}\approx2.5\times10^{18}\,\mathrm{cm}$, well outside the ejecta.

\textbf{SN~2017eaw ($+$2330~d).}
Pre-explosion HST and Spitzer imaging revealed a $\gtrsim 2\times10^{-5}~M_{\odot}$ CSM dust shell plus a $\sim$20\% rise in 4.5~$\mu$m flux in the final $\sim$3~yr before explosion \citep{Kilpatrick18}; progenitor analyses converge on $M_{\rm ZAMS}\approx 12$--$15~M_{\odot}$ \citep{VanDyk19,Rui19}. The CO first overtone was detected from $+$124 to $+$205~d with masses of $\sim 10^{-4}~M_{\odot}$ in LTE \citep{Rho18}, and \citet{Tinyanont19} reported a strong first-overtone feature at $+$389~d implying $\sim 10^{-3}~M_{\odot}$ of CO at $\sim$1800~K. JWST/MIRI epochs at $\sim$5.4~yr \citep{Shahbandeh23} and $\sim$6.4~yr \citep{Pearson2025} found a cool silicate component of $\sim 5.5\times10^{-4}~M_{\odot}$ at $T\approx 160$~K, with no significant evolution between cycles. At $+$2330~d, our MIRI SED shows the classic late-phase double-peaked structure: F560W $\approx 2.85~\mu$Jy rises to F1000W $\approx 37~\mu$Jy, dips through F1130W--F1500W, and peaks at F2100W $\approx 103~\mu$Jy. A two-temperature silicate fit yields $T_{\rm hot} \sim 1734$~K ($M_{\rm hot} \sim 5.30 \times10^{-8}~M_{\odot}$) and $T_{\rm cool} \sim 149 $~K with $M_{\rm cool} \sim 7.28 \times10^{-4}~M_{\odot}$, agreeing with \citet{Pearson2025}.

A second epoch of JWST observations of SN~2017eaw show no evolution in dust temperature or mass \citep{Pearson2025}. This stability disfavors freshly condensing grains, which would be expected to cool and grow over time, and instead points to a pre-existing dust heated by the fading supernova. Such pre-existing dust need not arise from only a distant light echo and may instead reside near the ejecta-CSM interaction region, provided it is not located significantly interior to the expanding ejecta. If the MIR emission is instead interpreted as a light echo, the dust must lie at $R_{\rm echo}\gtrsim c\times t\,(2330\,{\rm d})/2\approx3\times10^{18}$~cm, far outside the ejecta (the boxy H$\alpha$ profile gives $R_{\rm ej}\approx1.5\times10^{17}$~cm). For SN~2017eaw, the MIRI data favor a dominant pre-existing dust component, with any contribution from interaction-region dust likely modest and secondary. The largely unattenuated broad H$\alpha$ profile does not require newly formed ejecta dust, although the modest asymmetry noted by \citet{Pearson2025} may be consistent with some internal dust production.


 
\textbf{SN~2017gmr ($+$2335~d).}
\citet{17gmr_Andrews_Jen} reported a $\sim$500~$R_{\odot}$ progenitor radius and $M(^{56}\mathrm{Ni})$ up to $0.130\pm 0.026~M_{\odot}$, with an additional energy source for the first $\sim$2~d attributed to CSM interaction.  The earliest Keck HIRES spectra showed weak narrow H$\alpha$ emission but no other flash-ionization lines, and the H$\alpha$ profile also exhibited a weak ledge-like feature. \citet{Utrobin21_SN2017gmr} inferred an explosion energy of $\sim 10^{52}$~erg with bipolar $^{56}$Ni ejecta. At $+$2335~d, the MIRI SED is similar in shape to that of SN~2017eaw: F560W $\approx 2.4~\mu$Jy rises to F1000W $\approx 30~\mu$Jy and F2100W $\approx 47~\mu$Jy. A two-temperature silicate fit yields $T_{\rm hot} \sim 1580$~K ($M_{\rm hot} \sim 3.26 \times10^{-7}~M_{\odot}$) and $T_{\rm cool} \sim 174$~K with $M_{\rm cool} \sim 9.89 \times10^{-4}~M_{\odot}$.

For weak ongoing interaction, adopting $L_{\rm heat}\sim(1\text{--}10)\times10^{38}\,\mathrm{erg\,s^{-1}}$, the cool component ($T_{\rm d}=174\,\mathrm{K}$) implies $R_{\rm eq}\approx(6\text{--}20)\times10^{15}\,\mathrm{cm}$. This lies well within the ejecta radius inferred from the late-time Fe~II velocity ($v\approx3500\,\mathrm{km\,s^{-1}}$), corresponding to $R_{\rm ej}\approx7.1\times10^{16}\,\mathrm{cm}$, suggestive of dust in the ejecta or a CDS if weak interaction provides the heating. This interpretation is supported by the day +312 spectrum, which shows a multi-peaked H\(\alpha\) profile with red-wing attenuation, suggesting that some fraction of the dust responsible for the asymmetry may have formed newly in the ejecta \citep{17gmr_Andrews_Jen}. If the dust is heated by the earlier peak or plateau luminosity ($L\sim10^{42}\text{--}10^{43}\,\mathrm{erg\,s^{-1}}$), the same temperature requires $R_{\rm eq}\sim(0.6\text{--}2)\times10^{18}\,\mathrm{cm}$, placing the emitting grains well outside the ejecta and favoring pre-existing circumstellar or interstellar dust. The light travel scale at this epoch, $ct/2\sim3\times10^{18}\,\mathrm{cm}$, can also be consistent with an IR echo. In the absence of CO constraints, late-time spectroscopy, or a measured heating luminosity, the dust origin remains ambiguous.

\subsection{CO Detections in the Sample}
\label{sec:CO}
 
CO has been securely detected in only four of our 11 SNe (SN~2022acko, SN~2023ixf, SN~2024ggi and SN~2017eaw). For SN~2022acko, the first JWST spectrum at $+$50~d showed no CO and placed an upper limit of $<10^{-8}~M_{\odot}$ on the CO mass \citep{Shahbandeh24}, while CO was detected at later epochs. By $+$259~d, JWST medium-resolution spectra show weak CO first-overtone and fundamental-band emission emerging above the continuum, and these features persist at $+$368~d, although they remain weak and are not accompanied by clear SiO or dust signatures \citep{Medler2026}. The first overtone was also independently identified in ground-based NIR spectra at $+$289~d \citep{Medler_22acko_HISS}, consistent with relatively late CO formation in this object. In SN~2023ixf, ground-based NIR
spectroscopy first revealed the CO first overtone between $+$199 and $+$307~d
\citep{Park2025_SN2023ixf, Hsu2026_23ixf_nebular}, and the JWST subsequently
detected both the first overtone and fundamental bands at $+$250--720~d
\citep{Medler2025_SN2023ixf}. In SN~2024ggi, \citet{Mera2026_SN2024ggi}
detected CO at $+$285 and $+$385~d, following an earlier ground-based
detection at $+$250--319~d \citep{Hueichapan2026_SN2024ggi}; CO was absent at
$+$55~d \citep{Baron2025}, constraining the onset of CO formation in this
strongly interacting object to between $\sim$55 and $\sim$285~d. For SN~2017eaw, the CO first overtone was identified at $+$124--205~d \citep{Rho18} and subsequently
at $+$389--566~d \citep{Tinyanont19}. The remaining objects lack published NIR/MIR spectroscopy, so we cannot distinguish between true CO absence and observational incompleteness. The onset of CO emission among the observed objects is consistent with the expectation that CO formation in normal SNe~IIP begins between $\sim$100 and $\sim$300~d \citep{Sarangi_2013, Sarangi18}.

\vspace{-0.3em}
\subsection{Late-Time Optical Spectroscopy}
\label{sec:latetime_spectra}
 
The late time optical spectra span $+$638~d (SN~2022acko) to $+$2668~d (SN~2017eaw) and probe the nebular ejecta at epochs contemporaneous with or bracketing our MIRI observations. All five show H$\alpha$ emission, along with [\ion{O}{1}]~$\lambda\lambda$6300,\,6364 and [\ion{Ca}{2}]~$\lambda\lambda$7291,\,7323. The H$\alpha$ line profiles, shown in velocity space in the right panel of Fig.~\ref{fig:all_spectra_detection}, exhibit a diversity of morphologies that track the dust and CSM properties inferred from the MIRI SEDs. 
SN~2023ixf at $+$1002~d displays a broad and structured H$\alpha$ profile, extending to $\sim\pm10{,}000$~km~s$^{-1}$, with a pronounced blue-red asymmetry: the blueshifted wing is brighter than the redshifted wing \citep{Hsu2026_23ixf_nebular}. While SN~2017eaw may show marginally broader overall line wings, SN~2023ixf stands out for the complexity and asymmetry of its profile. This asymmetry is a classic signature of dust internal to the H$\alpha$-emitting region, which preferentially attenuates photons from the receding (redshifted) hemisphere \citep{Lucy89,Bevan2016_1987A, Dessart2025_dusty_typeII}. In that context, the profile is qualitatively consistent with the presence of a substantial cool-dust component $\sim 7\times10^{-3}~M_{\odot}$, inferred from the MIRI SED at a comparable epoch, although a dedicated line-profile analysis would be required for a quantitative comparison. SN~2024ggi at $+$671~d shows a similarly broad H$\alpha$ profile with a subtle blue-red asymmetry, consistent with the onset of dust formation indicated by its MIRI SED at $+$660~d.

SN~2017eaw at $+$2668~d exhibits a distinctive boxy, flat-topped H$\alpha$ profile that first emerged around $+$900~d \citep{Weil2020} and has persisted for over five years \citep{Pearson2025}.  Radiative transfer models \citep{Dessart2025_dusty_typeII} attribute this morphology to dust at the ejecta--CSM interface, with a predicted mass of several $\times10^{-4}~M_{\odot}$, consistent with our derived value of
$7.28\times10^{-4}~M_{\odot}$ at $+2330$~d; however, the symmetric profile suggests limited line-of-sight dust extinction. These models indicate that dust in the CDS dominates the line-profile asymmetries over ejecta dust at $\lesssim1000$ d, although whether this remains true at the later epochs probed here is untested. SN~2020jfo at $+1376$~d shows weaker emission lines and a narrower H$\alpha$ profile, consistent with a less massive ejecta \citep{Sollerman2021_SN2020jfo}, while the faint SN~2022acko exhibits minimal line emission, consistent with its low luminosity and $^{56}$Ni mass. The late-time spectra broadly support the MIRI
dust-mass estimates, the dust rich SNe (SN~2023ixf and SN~2024ggi) show the strongest internal-dust signatures, whereas like SN~2022acko, SN~2020jfo lacks significant line asymmetries. SN~2017eaw is a caveat to this picture; despite one of the lowest inferred dust masses in the sample, its persistent boxy H$\alpha$ traces sustained CSM interaction.

\subsection{Sample Wide Trends and Properties}
\label{sec:trends}

\subsubsection{Warm and Cool Dust Components}
\label{sec:warm_cool}

Every SN in the sample requires at least two components to reproduce the MIRI SEDs (see Table~\ref{tab:fit_params}). The warm/hot component is often sampled by only a single MIRI data point in many SNe and is therefore poorly constrained. We therefore refrain from interpreting its physical properties or origin. SN~2017eaw provides a useful reference, as its warm excess has been identified in multiple JWST/MIRI epochs and is supported by extensive NIR coverage \citep{Shahbandeh23, Pearson2025}. For the remaining objects without NIR coverage, however, the available MIRI data only indicate the presence of an additional warm component.

SN~2023ixf required a three component fit: hot graphite at $985$~K ($1.15\times10^{-6}~M_{\odot}$), cool graphite at $278$~K ($6.05\times10^{-3}~M_{\odot}$), and cold silicate at $222$~K ($1.30\times10^{-3}~M_{\odot}$). The $+$250--720~d MIRI spectra of \citet{Medler2025_SN2023ixf} show that the MIR SED is continuum dominated at these phases, with little to no emission-line contamination, so our photometry likely traces the continuum. \citet{Singh2026_23ixf} disentangled the early IR echo from newly forming internal dust and inferred a growing silicate component reaching $\sim2\times10^{-3}~M_{\odot}$ by $+$750~d, so the $\approx7.4\times10^{-3}~M_{\odot}$ of cool dust we infer at $+$992~d suggests continued condensation between the two epochs. Contemporaneous NIR spectroscopy reveals rapid molecular evolution in SN~2024ggi, with the CO-emitting gas cooling, becoming increasingly clumped, and its CO mass decreasing by nearly an order of magnitude between $+285$ and $+385$~d \citep{Mera2026_SN2024ggi}. These changes are consistent with the subsequent detection of $1.44\times10^{-3}~M_{\odot}$ of silicate dust in our MIRI observations $\sim300$~d later.

\begin{figure*}[htp]
 	\centering
 \includegraphics[width = \textwidth]{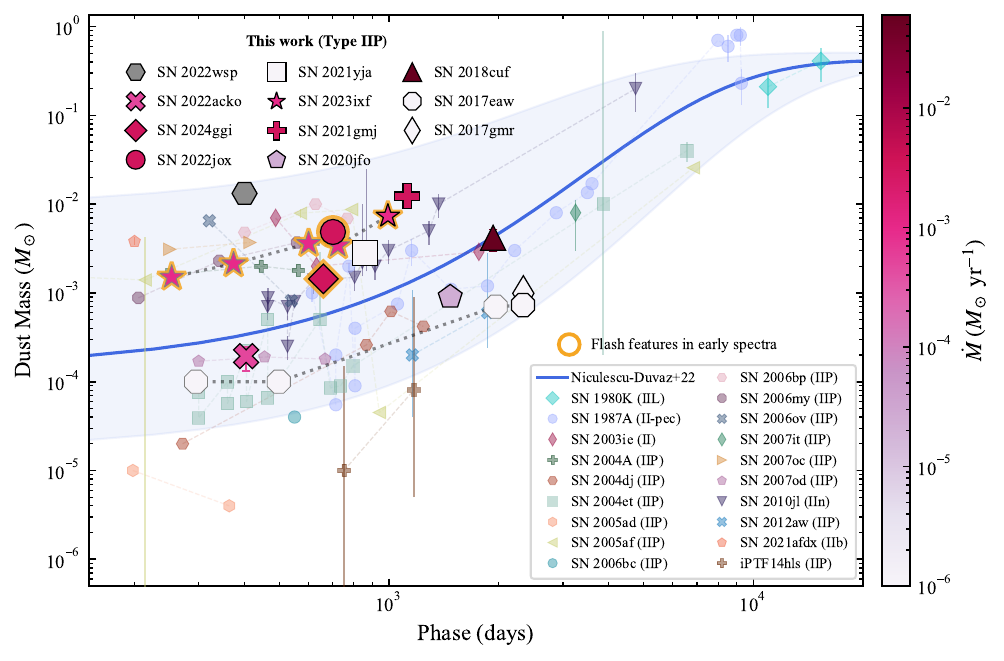}
 \vspace{-10pt}
 \caption{Dust mass as a function of phase for our 11 SNe (large colored symbols) and comparison objects from the literature (small symbols). Symbol color for only the 11 SNe encodes the progenitor mass-loss rate $\dot{M}$, as shown in the color-bar, with gold rings marking SNe with detected early flash-ionization features. The mass-loss rates shown in this figure are heterogeneous literature estimates derived under different assumptions and should be viewed as qualitative indicators of relative CSM strength.}
     \label{fig:dust_timeline_mdot}
\end{figure*}

We note that in all cases, the cool dust component dominates the total dust mass budget. Silicate is detected in all 11 objects, making it the prevalent condensate in our sample, consistent with oxygen-rich ejecta compositions expected for RSG progenitors in the $\sim$9-20~$M_{\odot}$ range \citep{Sarangi18} and with the clear 10~$\mu$m silicate emission feature present in the MIRI photometry of most objects. For 10 of the 11 SNe, silicate is the sole cool-dust species; SN~2023ixf is the exception, requiring an additional graphite component at $278$~K ($6.05\times10^{-3}~M_{\odot}$) that dominates the cool-dust mass budget, while the silicate at $222$~K ($1.30\times10^{-3}~M_{\odot}$) is actually the coldest of the three components. The coexistence of carbon and oxygen bearing grains at different temperatures may reflect condensation in chemically distinct zones of the ejecta, graphite forming in the C/O rich layers and silicate in the O/Si/Mg rich zones that may be more extensively exposed in this luminous, strongly interacting SN \citep{Fedkin2010}.
 
Cool dust temperatures in our sample range from $\sim$ 100 - 250 K, and the corresponding masses span two orders of magnitude from $\sim 2\times10^{-4}$ to $\sim 1.2\times10^{-2}~M_{\odot}$. These values bracket the JWST-derived dust masses reported for other Type~II SNe: SN~2004et at $+$18~yr harbors $\gtrsim 0.014~M_{\odot}$ of cool dust at $\sim$140~K \citep{Shahbandeh23}, while SN~1993J at $\sim$30~yr shows $\sim 3\times10^{-3}~M_{\odot}$ at $\sim$120~K \citep{Zsiros22, Szalai2025_SN1993J}. We emphasize that the MIRI derived masses should be regarded as lower limits on the total dust content, as dust colder than $\sim$80-100~K would radiate predominantly beyond 25~$\mu$m and remain inaccessible to these observations.

\subsubsection{Dust Mass Growth and the Influence of CSM}
\label{sec:dust_timeline}

To place our results in a broader CCSN context, we compare the 11 dust masses derived from our JWST/MIRI fits with values compiled from the literature in Fig.~\ref{fig:dust_timeline_mdot}, including early Spitzer-based estimates \citep{Fox2011, Szalai13, Szalai19}, late-time optical line-profile studies with \textsc{damocles} \citep{Bevan2016_1987A, Niculescu-Duvaz2022}, and far-IR/submillimeter measurements of SN~1987A \citep{Matsuura11, Indebetouw14, Matsuura2015}. These compilations trace dust growth on decade timescales, and we show the population model of \citet{Niculescu-Duvaz2022} for reference. Because the literature masses are heterogeneous in method and underlying assumptions, they should be viewed as benchmarks rather than homogeneous measurements. Our MIRI-based analysis provides a uniform 5.6--25.5~\(\mu\)m dataset and common fitting framework, with sensitivity to cool dust inaccessible to earlier Spitzer studies.

The comparison of our JWST/MIRI sample with previous literature studies show that the overall broad picture has not changed. Even with sensitivity to colder dust, the masses we infer at $\sim$400--2300~d occupy the same general range ($\sim 10^{-4}$--$10^{-2}~M_{\odot}$) as previous estimates at comparable phases \citep{Niculescu-Duvaz2022}. A notable feature of Fig.~\ref{fig:dust_timeline_mdot} that emerges is the wide dispersion in dust mass at fixed phase among CCSNe. This dispersion is already apparent at the earliest
epochs, where SN~2022acko and SN~2022wsp occupy opposite ends of the
distribution at comparable phases. The
diversity persists at intermediate phases, where SN~2024ggi, SN~2022jox,
and SN~2021yja span a wide range in inferred dust masses, with
SN~2022jox at the upper end of this subset. By $\sim$1000~d, SN~2023ixf
remains among the most dust rich objects at comparable epochs, and
between $\sim$1000 and 1500~d SN~2021gmj is much dustier than
SN~2020jfo. At the very late phases, $\sim$1900--2300~d, SN~2018cuf is among the more dust rich objects in our sample as compared to SN~2017eaw and SN~2017gmr, and falls within the range suggested by the \citet{Niculescu-Duvaz2022} model.

We note that Fig.~\ref{fig:dust_timeline_mdot} suggests that a source of the scatter might be linked to the CSM material ejected in the final years to decades before core collapse that are now recognized around a substantial fraction ($\gtrsim 60$\%) of SN~II progenitors \citep{Bruch2021, Bruch2023, Irani2024}. This trend is visible in a relative comparison between objects with clear CSM signatures and those without strong evidence for dense CSM. SNe with strong evidence for dense CSM interaction, corresponding to $\dot{M} \gtrsim 10^{-3}~M_{\odot},\mathrm{yr}^{-1}$, such as SN~2023ixf, SN~2024ggi, and SN~2022jox, have a median cool-dust mass of $\sim 5\times10^{-3}~M_{\odot}$. In contrast, objects without strong CSM evidence, including SN~2022acko, SN~2020jfo, and SN~2017gmr, have a median cool-dust mass of only $\sim 5\times10^{-4}~M_{\odot}$ at comparable phases. SN~2021yja also falls within this broad dust-mass range, although its progenitor mass-loss rate is inferred to be relatively weak and remains debated so its placement relative to the CSM rich subset is less secure \citep{Hosseinzadeh2022, Vasylyev2024_SN2021yja, Kozyreva2022_SN2021yja}. Even after accounting for the overall phase dependence of dust growth, the CSM rich objects remain elevated by roughly an order of magnitude, although this trend should be regarded as a weak indirect correlation given the heterogeneous mass-loss estimates in the literature derived under varying assumptions and the current limited sample size.

\begin{figure}[htp]
	\centering
\includegraphics[width = \columnwidth]{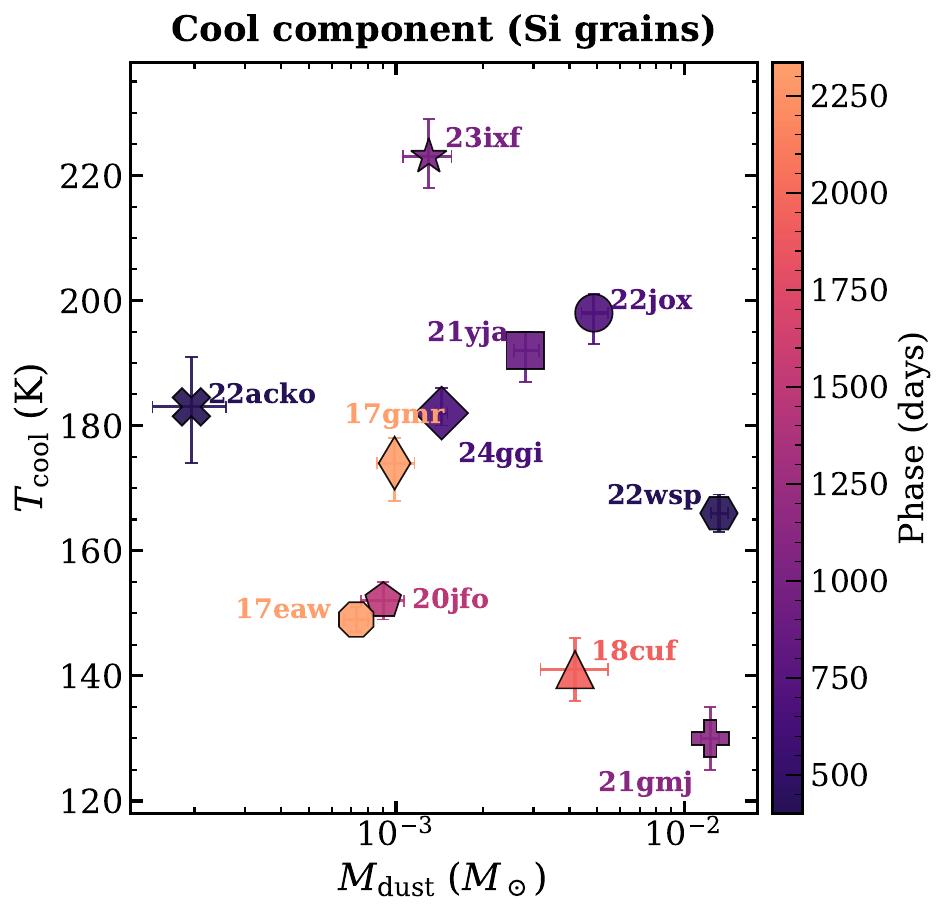}
\vspace{-2em}
    \caption{Cool component dust temperature versus dust mass for the 11~SNe in the sample, color-coded by the phase. }
    \label{fig:Tcool_Mdust}
    \vspace{-2em}
\end{figure}

The enhanced dust emission may be associated with ejecta-CSM interaction, which can create a dense, rapidly cooling environment favorable for dust formation. One physical mechanism is condensation in the CDS formed between the forward and reverse shocks during continued interaction with extended CSM. In this context, \citet{Sarangi_Slavin2022} showed that post-shock gas in the shell can cool to conditions for dust nucleation, and their models support dust masses of $10^{-3}$ to $0.8\,M_{\odot}$ depending on the mass-loss rate and ejecta properties. This range is broad, however, and overlaps much of the observed CCSN dust-mass distribution. Observational studies of interacting SNe such as SN~1998S \citep{Pozzo04, Mauerhan_Smith2012}, SN~2005ip \citep{Fox2010_05ip, Smith09}, and SN~2010jl \citep{Gall2014, Smith2012, Smith2026_SN2010jl} have likewise been interpreted in terms of dust formation in a post-shock dense environment. The radiative-transfer calculations of \citet{Dessart2025_dusty_typeII} are broadly consistent with this picture from the optical side, showing that dust in the CDS can produce broad, asymmetric H$\alpha$ profiles at late times.

A second, related possibility is that some of the dust emission arises in cooled material associated with the innermost circumstellar environment that is overrun by the ejecta shortly after explosion. This left over material in the central regions may later contribute to the dust-emitting region, but it is not yet clear whether it should be regarded as a distinct dust reservoir from the CDS or simply as part of the same evolving post-shock structure. Recent simulations by \citet{Takei2025} begin to address this regime by extending the confined-CSM framework to compact shells ejected shortly before core collapse. In those models, the ejecta sweep up $\sim 0.01$--$0.1~M_{\odot}$ of confined CSM, and the resulting dense shell cools and forms dust on timescales of order $\sim$200--900~d. The dust yield in those calculations is dominated by the shocked ejecta component of the shell, while the role of the swept-up circumstellar gas itself remains poorly constrained. They discuss that a cold dense shell in confined CSM can also form molecules, including CO, though the faster cooling likely leads to different chemistry than in freely expanding ejecta.

Taken together, these scenarios are qualitatively consistent with our finding that SN~2023ixf, SN~2024ggi, and SN~2022jox combine relatively large dust masses with evidence for early, short-lived flash features, while SN~2021gmj may also fall into this category based on its early light-curve behavior and ledge feature \citep{Bostroem2024_SN2023ixf, Shrestha2024,22jox_Andrews2024, 21gmj_Meza_Retamal_2024}. The spectra of these SNe point to dense CSM in the immediate progenitor environment, which may create conditions favorable for enhanced dust formation over low density quiescent steady winds. The geometry of the CSM may also influence these observations. For the same total CSM mass, a confined or asymmetric distribution can produce higher local densities and stronger narrow emission features than a more extended, diffuse distribution. We therefore view the link between early flash signatures, confined CSM, and later dust yield as tentative and indirect, likely depending on the CSM geometry, density structure, and subsequent cooling efficiency, and requiring further study for confirmation. Future dust formation models may benefit from incorporating confined CSM geometries and densities, alongside comparison with time-resolved JWST observations, to examine whether such environments could truly affect dust in these systems.

\begin{figure*}[htp]
 	\centering
 \includegraphics[width = \linewidth]{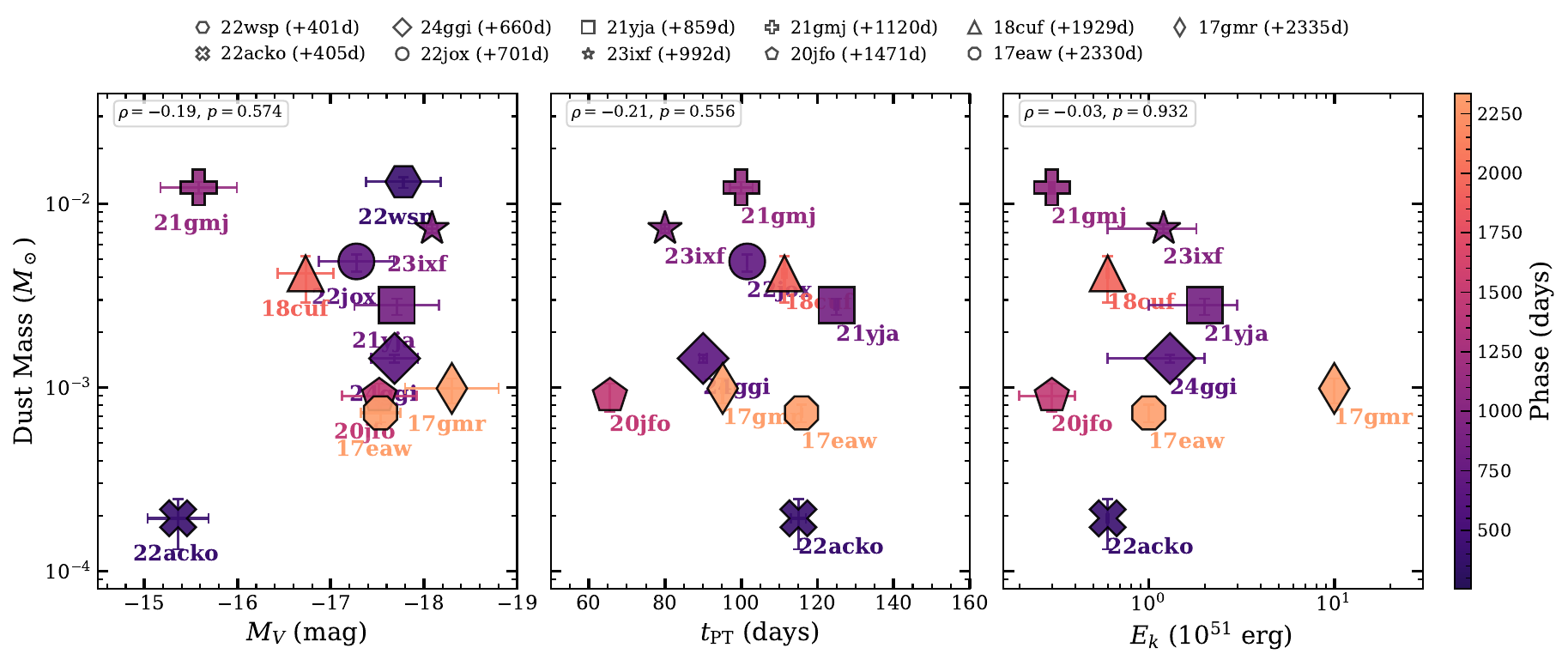}
  \vspace{-10pt}
  \caption{Dust mass versus peak absolute magnitude ($M_V$), plateau duration ($t_{\rm PT}$), and explosion energy ($E_k$) for the SNe in the sample, with points color-coded by phase since explosion. Spearman rank-correlation coefficients and $p$-values are shown; no significant correlations are found with these intrinsic SN properties.}
     \label{fig:dust_mass_no_correlation}
\end{figure*}

\vspace{-0.5em}
\subsubsection{Dust Temperature and Mass Evolution}
\label{sec:temperature_evolution}

Figure~\ref{fig:Tcool_Mdust} shows the cool component dust temperature $T_{\rm cool}$ plotted against dust mass, with points color-coded by the phase relative to the explosion epoch. The poorly constrained hot components are excluded from this analysis. One weak tendency emerges in Figure~\ref{fig:Tcool_Mdust}: objects observed at earlier phases (dark/purple symbols) cluster at higher temperatures ($\sim$180--200~K) and lower masses ($\sim 10^{-4}$--$10^{-3}~M_{\odot}$), while objects at later phases (orange/yellow symbols) have cooled to $\sim$130--175~K with masses reaching $\sim 10^{-3}$--$10^{-2}~M_{\odot}$. The exception is SN 2021gmj, which is discussed more below. This inverse correlation between temperature and phase is the expected signature of dust that cools as the ejecta expand and the main heating sources (radioactive decay and CSM interaction) decline, as originally predicted by one-zone models of supernova dust formation and evolution \citep{Kozma_1998, Sarangi18}. Figure~\ref{fig:Tcool_Mdust} also shows scatter in the sample between $T_{\rm cool}$ and $M_{\rm dust}$, in line with a Spearman rank test that finds no significant correlation ($\rho=-0.14$, $p=0.69$). For optically thin modified-blackbody emission, the inferred dust mass scales approximately as $M_{\rm dust}\propto T^{-(4+\beta)}$, where $\beta$ is the dust emissivity index \citep{Draine2003,Draine07}. Thus, even modest uncertainties in the cool-component temperature can translate into large changes in the inferred dust mass, particularly at low temperatures \citep{Matsuura11,Gall2014}. SN~2021gmj, which has the lowest fitted temperature in the sample ($T_{\rm cool}=130 \pm 5$~K, statistical uncertainty) correspondingly has the largest inferred dust mass ($1.23\times10^{-2}~M_{\odot}$). This $T$--$M$ degeneracy is a current limitation of single epoch SED fitting.

\vspace{-1 em}
\subsubsection{Dust Mass vs. Intrinsic SN Explosion Properties}

We examine whether the inferred dust mass is correlated with SN properties by comparing it with the peak absolute magnitude ($M_V$), plateau duration ($t_{\rm PT}$), and explosion energy ($E_{\rm k}$), adopting values from the literature references associated with each object in Table~\ref{tab:sample}, as shown in Figure~\ref{fig:dust_mass_no_correlation}. We do not detect a statistically significant monotonic relation with any of these quantities. Spearman rank correlation tests yield weak positive correlations with $M_V$ ($\rho = 0.19$, $p = 0.57$) and $t_{\rm PT}$ ($\rho = 0.21$, $p = 0.56$), and essentially no correlation with $E_{\rm k}$ ($\rho = 0.03$, $p = 0.93$). In all cases, the large $p$-values indicate there are no trends in the data. 

This absence of a simple trend is consistent with the broader theoretical and observational picture. Dust formation models predict that the dust yield depends on local ejecta conditions including the density and temperature evolution, chemical stratification, radioactive heating, mixing, clumping, and grain-growth timescales \citep[e.g.,][]{Todini01,Nozawa2003,Bianchi2007,Sarangi_2013,Sarangi2014}. Calculations that explicitly vary explosion properties further show that progenitor mass, metallicity, explosion energy, and explosion engine can affect the onset of dust formation, grain sizes, and dust composition \citep{Marassi2019, Brooker2022}. However, these effects are coupled and may not necessarily produce a monotonic relation. Although explosion energy directly affects the ejecta density and thermal evolution, its impact on dust production is modulated by other properties such as progenitor structure, mixing, chemical composition, and heating history. For instance, SN~2017gmr and SN~2017eaw differ substantially in explosion energy and $^{56}$Ni yield, yet their inferred dust masses at similar late phases are comparable within the uncertainties.

\subsection{Dust masses across CCSN subtypes}\label{app:dust_by_type}

\begin{figure*}[htp]
 	\centering
 \includegraphics[width = 0.8\textwidth]{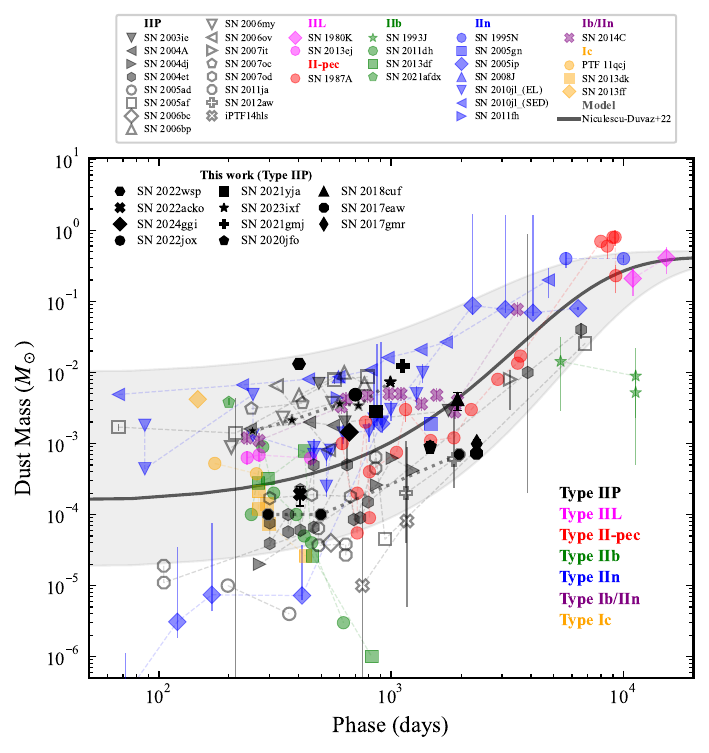}
  \vspace{-10pt}
  \caption{Compilation of dust masses as a function of phase for CCSNe spanning Type~IIP, IIL, II-pec, IIb, IIn, Ib/IIn, and Ic events. Symbols denote individual objects, with colors indicating subtype and the black curve showing the empirical trend from \citet{Niculescu-Duvaz2022}. For SN~2010jl, dust masses derived from emission-line (EL) modeling and from near- to mid-infrared SED fitting differ substantially, underscoring the effects of heterogeneous techniques used for estimating dust masses. The data for this compilation plot is collectively gathered from \citet{Zsiros22, Niculescu-Duvaz2022, Fox2020_SN2005ip, Szalai19_spitzer, Hosseinzadeh23, Zsiros24, Tinyanont2025_SN2014C, Clayton2025_SN1995N, Smith2026_SN2010jl} and references within.}
     \label{fig:dust_timeline_bytype}
\end{figure*}

Figure~\ref{fig:dust_timeline_bytype} places our Type IIP sample in the broader context of dust masses measured across the CCSN population. At phases $\lesssim10^3$~d, the inferred dust masses of Type~IIP/IIL, IIb, II-pec, and IIn events largely overlap, with no clear subtype-dependent separation in the available compilation. Nevertheless, a small number of Type~IIn events preferentially occupy the upper end of the distribution, consistent with the expectation that dense CSM environments may provide favorable conditions for dust formation, grain survival, or the heating of pre-existing circumstellar dust \citep{Fox2011,Gall11,Sarangi18}. At later epochs, Type~IIn SNe increasingly
populate the upper envelope of the compiled sample; however, this apparent
enhancement is based on a limited number of objects and is likely shaped by
the strong observational preference for IR bright, long lived
interaction powered transients, as well as by differences in dust mass
inference methods \citep{Fox2011, Szalai13, Smith2012, Gall2014, Fox2020_SN2005ip}. Large dust masses at very late phases are not exclusive to
interacting events: SN~1987A and SN~1980K (classified as Type~II-pec and
Type~IIL, respectively) reach dust masses comparable to the most dust-rich
Type~IIn SNe at late times suggesting that substantial dust reservoirs may also arise in non-interacting explosions \citep{Matsuura11,Bevan2016_1987A}. The late-time Type~IIb population remains particularly unconstrained, with
SN~1993J providing one of the few available long baseline constraints and
showing comparatively modest dust growth. The stripped envelope SNe Type Ib/Ic population remains the most undersampled among the subtypes. We therefore interpret the apparent late time excess of Type~IIn dust masses to be treated with caution as it may reflect intrinsic differences in dust formation and processing in dense CSM environments, observational selection toward the most favorable systems, or systematic differences among analysis techniques. A larger, uniformly selected sample with consistent IR diagnostics will be required to determine whether CCSN progenitors across different subtypes is fundamentally linked to the efficiency of dust production.

\vspace{-1em}
\subsection{Implications to Dust Budget in Early Universe}
\label{sec:dust_budget}

A central motivation for measuring dust in CCSNe is the large dust masses, $\sim10^{7}$--$10^{8}~M_{\odot}$, inferred for galaxies at $z>6$, when the Universe was too young for AGB stars to contribute significantly and CCSNe are thought to be the primary stellar dust source \citep{Dwek2011,Gall2011,Marrone18,Nanni2025}. This argument applies to the full CCSN population: hydrogen-rich Type~IIP/L events constitute the majority ($\sim 70\%$), stripped-envelope Type~IIb/Ib/Ic explosions account for most of the remainder, and strongly interacting Type~IIn SNe occur only at the few-percent level \citep{Li2011,Smith2011,Shivvers2017}; consequently, any population-averaged dust yield is anchored primarily by the normal Type~II-P events that dominate the rate. 

A simple rate argument illustrates the required dust yield. For a rigorously star-forming galaxy with ${\rm SFR}\approx100~M_{\odot}\,{\rm yr^{-1}}$ and one CCSN per $\sim100~M_{\odot}$ of stars formed \citep{Botticella2012, Madau2014}, the CCSN rate is $\sim1~{\rm yr^{-1}}$, corresponding to $\sim5\times10^{8}$ explosions over the $\sim500$~Myr available by $z\sim6$. Producing $10^{8}~M_{\odot}$ of dust therefore requires a surviving yield of $\sim0.2~M_{\odot}$ per SN. If the observed dust is predominantly formed in the ejecta, reverse shocks are expected to destroy $\sim10$--$90\%$ of it  \citep{Gall2011,Slavin2020,Kirchschlager2019,Kirchschlager2022}, implying initial dust masses of $\sim0.2$--$2~M_{\odot}$ per event. This calculation becomes less direct if a significant fraction of the observed dust instead resides in the CDS or pre-existing CSM. In the most extreme high-$z$ starbursts with ${\rm SFR}\sim2000$--$3000~M_{\odot}\,{\rm yr^{-1}}$, the required surviving dust yield drops to only $\sim0.007$--$0.03~M_{\odot}$ per CCSN. For ${\rm SFR}\approx3000~M_{\odot}\,{\rm yr^{-1}}$ and star-formation durations of 100--500~Myr, this implies $N_{\rm CCSN}\approx 10^{9}-10^{10}$ and the same required surviving yield of $\sim0.007$--$0.03~M_{\odot}$ per event. The lower end of this range is set by the dust-rich SNe in our sample, and even the median falls short by only a factor of $\sim$2 under the most optimistic assumptions. For a typical star-forming galaxy, however, the shortfall remains substantial. The fitted cool-dust masses in our Type IIP sample range from $\sim2\times10^{-4}~M_{\odot}$ for SN~2022acko to $\sim1.2\times10^{-2}~M_{\odot}$ for SN~2021gmj, with a median of $\sim3\times10^{-3}~M_{\odot}$, well below even the relaxed seed
requirement of $\sim0.01$--$0.1~M_{\odot}$ for scenarios in which most of the dust mass grows later in the SN ejecta or ISM \citep{Draine2009, Asano2013, Ferrara2016}. At this median yield, $5\times10^{8}$ SNe would produce only $\sim1.5\times10^{6}~M_{\odot}$ of dust over 500~Myr, about a factor of 70 below a $10^{8}~M_{\odot}$ dust budget, and only the strongest-CSM objects reach the lower end of the grain-growth seed requirement at 1--6~yr after explosion. 

The diversity of CCSN subtypes is important when interpreting the population-averaged dust yield. While Type~IIn SNe tend to occupy the upper end of the late-time ($\gtrsim2000$~d) dust-mass distribution (Figure~\ref{fig:dust_timeline_bytype}), their dense CSM may enhance dust formation or survival, or simply increase the detectability of pre-existing circumstellar dust \citep{Fox2011,Gall11,Sarangi18, Clayton2025_SN1995N, Smith2026_SN2010jl}. Although Type~IIn SNe comprise only $\sim$5--10\% of the CCSN population, their potentially larger dust yields could contribute disproportionately to the total dust budget if representative, but they remain too rare to dominate it. Conversely, hydrogen-rich Type~II-P SNe constitute the majority of CCSNe, yet remain sparsely sampled beyond a few years after explosion, making their characteristic late-time dust yield poorly constrained. SESNe account for a further $\sim$25$-$30\% of the CCSN rate; although their carbon- and oxygen-rich ejecta may favor different dust chemistry, the limited late-time observations provide no compelling evidence that they systematically produce larger dust masses than normal Type~II SNe. The highest dust yields observed among rare or observationally biased CCSN subtypes should therefore not be regarded as representative of the population as a whole.

We also emphasize that the inferred dust masses are derived under the assumption of optically thin emission and therefore trace only the dust directly visible to MIRI. Any dust residing within the compact inner ejecta may remain undetected. Theoretical models suggest that dust formation in this region continues beyond the epochs probed here, and even relatively small dust masses at earlier times can become optically thick. Observations of SN~1987A demonstrate that substantial quantities of interior dust may only become apparent several years after explosion and at longer wavelengths with facilities such as ALMA \citep[e.g.,][]{Matsuura11,Matsuura2015}.
Following \citet{Sarangi2025_SN2005af}, the optical depth at $10~\mu{\rm m}$ is given by $\tau(10~\mu{\rm m}) = 3M_{\rm d}\kappa/[4\pi(vt)^2]$, where $\kappa_{\rm Si}(10~\mu{\rm m})\approx3400~{\rm cm^{2}~g^{-1}}$ \citep{Draine2007} and $v\approx1000$--$1400~{\rm km~s^{-1}}$ is representative of the inner ejecta. For SNe observed at $\lesssim 1000$~d, even modest dust masses confined to this compact region can yield $\tau \gg 1$, effectively obscuring a significant fraction of the dust from direct detection. In this optically thick regime, the emergent MIR flux becomes insensitive to the total dust mass and instead depends primarily on the emitting surface area and temperature, scaling approximately as $F_{\nu}\approx\pi B_{\nu}(T)(vt/d)^2$ \citep{Dwek2015}. For representative parameters ($v = 1200~{\rm km~s^{-1}}$, $t = 1000$~d, and $d = 7$~Mpc), this yields a flux of $\sim 30~\mu{\rm Jy}$ at $21~\mu{\rm m}$ for $T = 150$~K, but only a few $\mu{\rm Jy}$ for $T \lesssim 100$~K. Consequently, a substantial hidden dust reservoir would likely be relatively cold, and thus increasingly difficult to detect at earlier epochs. Such emission becomes more accessible at later times ($\gtrsim 1500$~d), when the ejecta have expanded and cooled sufficiently.

Additional dust reservoirs may also lie outside the direct sensitivity of the MIRI data. In the \citet{Sarangi2025_SN2005af} chemical-kinetic model, amorphous carbon accounts for \(\sim80\%\) of the total dust mass, reaching \(\sim0.03~M_{\odot}\) by \(\gtrsim3000\)~d. Because amorphous carbon is largely featureless in the MIR, a cold carbon component would be difficult to distinguish from the broadband thermal emission. Likewise, very cold dust is poorly constrained by MIRI. The \(\sim0.5~M_{\odot}\) of \(\sim20\)~K dust inferred in SN~1987A radiates primarily at far-infrared and submillimeter wavelengths beyond the MIRI bandpass \citep{Matsuura2015}. A more direct comparison would be to calculate the MIRI fluxes predicted by the chemical kinetic models at the epochs of our observations and compare them with the observed SEDs. This would test whether the predicted \(\sim0.03~M_{\odot}\) of dust is consistent with the data or whether most of it must remain undetectable by MIRI.

The resulting picture is that dust formation is a universal outcome of CCSNe: all 11 objects in our sample harbor detectable dust; but that the \emph{observable} MIR masses at 1--6~yr cannot, by themselves, settle the budget question. If the true per-event yield approaches the $\sim$0.003~$M_{\odot}$ of chemical model predictions, the population produces $\sim1.5\times10^{6}~M_{\odot}$ per galaxy over 500~Myr, still several times short before reverse-shock losses. Matching the full budget without ISM grain growth requires SN~1987A-like masses ($\sim$0.5~$M_{\odot}$) to be typical rather than exceptional and reverse-shock survival to be efficient; and the factor of $\sim$60 spread in dust mass at fixed phase within our sample, together with the dominance of low-mass progenitors in the CCSN rate, argues against adopting the most productive events as representative. The more reasonable conclusion is that CCSNe supply dust seeds at the $\sim10^{-3}$--$10^{-2}~M_{\odot}$ level comfortably within our measured range for the CSM-enhanced SNe, with subsequent growth in the ejecta as well as the ISM \citep{Draine2009, Asano2013, Ferrara2016}. 

\vspace{-1em}
\section{Conclusions}
\label{sec:conclusion}
 
We have presented JWST/MIRI photometry and SED dust modeling for a sample of 11 Type~IIP core-collapse supernovae, spanning rest-frame phases of $+$401 to $+$2335~d post-explosion. For every SN, we fit optically thin dust models (silicate, amorphous carbon, graphite), supplemented where required by a blackbody component to capture residual photospheric or warm continuum emission, using a common nine-band MCMC framework spanning 5.6--25.5~$\mu$m. Our principal conclusions are as follows.

\begin{enumerate}[leftmargin=0.5cm, nosep]
    \item \textbf{Dust is universally present in CCSNe.} All 11 SNe are detected in at least seven MIRI filters. No object in the sample is consistent with a dust free SED at the observed epoch, establishing warm to cool dust emission as a universal outcome in CCSNe on $\sim$1--6~yr timescales.
    
    \item \textbf{The SEDs show a coherent MIR time evolution.} The emission evolves from hot, featureless continua at $\sim$400~d to prominent 10~$\mu$m silicate emission by $\sim$700--900~d, and to cool, 15--25~$\mu$m silicate-dominated SEDs beyond $\sim$1000~d, consistent with dust growth and cooling.
    
    \item \textbf{Cool dust temperatures span 100--250~K and masses span two orders of magnitude across the sample.} Cool-component temperatures range from 100--250~K and cool-dust masses from $\sim2\times10^{-4}$ to $1.2\times10^{-2}~M_{\odot}$ in our sample, bridging the poorly explored $\sim$1--7 yr epoch of dust evolution. The inferred masses are lower limits owing to the limited sensitivity of MIRI to colder dust.
    
    \item \textbf{Dust mass shows no correlation with SN explosion properties.} Cool-dust mass does not show a significant relation with peak absolute magnitude $M_V$, plateau duration $t_{\rm PT}$, or explosion energy $E_k$. While explosion energy is expected to influence dust formation, its effects are likely coupled with other physical processes.
    
    \item \textbf{Pre-explosion mass loss may contribute to the dust mass diversity at fixed phase.} SNe with confined CSM signatures in their light curve or early spectra (SN~2023ixf, SN~2024ggi, SN~2021gmj, and SN~2022jox) tend to occupy the upper end of the cool-dust mass distribution at comparable phases, with a median of $\sim5\times10^{-3}~M_{\odot}$. This suggests that enhanced pre-explosion mass loss and the resulting CSM environment may influence the late time dust yield, although the correlation remains tentative.
    
    \item \textbf{CCSNe dust yields at 1-6 yr support seed dust formation but cannot by themselves account for high-z dust reservoirs.} The observed 1--6 yr CCSN dust masses may represent an early seed population, but they are unlikely to fully account for high-$z$ dust reservoirs unless substantial subsequent grain growth occurs. \\
\end{enumerate}

\noindent Several pathways can advance CCSN dust studies further than this present study. The most immediate is to move beyond single-epoch SEDs to multi-epoch JWST/MIRI imaging which can track whether the dust cools and grows in mass or merely fades with the declining radiation field, turning the current temperature mass degeneracy into a diagnostic of grain growth and revealing whether the apparent dust-mass plateau beyond $\sim$1500~d reflects genuine saturation or slower growth unresolved by our present baseline. A more complete census will require JWST/MIRI MRS spectroscopy to separate the dust continuum from line emission and to resolve the 5--28~$\mu$m emission into dust continuum, line cooling, and the silicate feature profile that constrains grain composition.

Finally, larger and more homogeneous samples ($\gtrsim30$ CCSNe with uniform phase sampling, contemporaneous CSM diagnostics, early NIR spectroscopy, and a broader representation of objects with confined CSM interaction) extended to baselines of $\gtrsim5$--10~yr would provide the statistical leverage to disentangle pre-explosion mass loss, progenitor properties, explosion parameters, and CSM environment as drivers of dust yield, while anchoring intermediate-age CCSNe to the substantial reservoirs seen in SN~1987A and young remnants. Together, early CO observations and long-term MIRI monitoring offer a route to mapping the full life cycle of dust in CCSNe.

\vspace{-2em}
\begin{acknowledgments}

Thank you to C. DeCoursey and J. Pierel for their help with JWST photometric reduction using space\textunderscore phot. This work is based in part on observations made with the NASA/ESA/CSA James Webb Space Telescope. The data were obtained from {MAST} at the Space Telescope Science Institute, which is operated by the Association of Universities for Research in Astronomy, Inc., under NASA contract NAS 5-03127 for JWST. These observations are associated with program GO3295 and GO7881. The complete set o JWST data presented in this article were obtained from the Mikulski Archive for Space Telescopes (MAST) at the Space Telescope Science Institute. The specific observations analyzed can be accessed via \dataset[doi:10.17909/fxr9-ck41
]{https://doi.org/10.17909/fxr9-ck41}

This work is also based on observations obtained at the international Gemini Observatory, a program of NSF's NOIRLab, which is managed by the Association of Universities for Research in Astronomy (AURA) under a cooperative agreement with the National Science Foundation. On behalf of the Gemini Observatory partnership: the National Science Foundation (United States), National Research Council (Canada), Agencia Nacional de Investigaci\'{o}n y Desarrollo (Chile), Ministerio de Ciencia, Tecnolog\'{i}a e Innovaci\'{o}n (Argentina), Minist\'{e}rio da Ci\^{e}ncia, Tecnologia, Inova\c{c}\~{o}es e Comunica\c{c}\~{o}es (Brazil), and Korea Astronomy and Space Science Institute (Republic of Korea). 

The NIR JHK imaging observations presented here were obtained with MMIRS at the MMT Observatory, a joint facility of the University of Arizona and the Smithsonian Institution.

This work was supported by a NASA Keck PI Data Award, administered by the NASA Exoplanet Science Institute. Data presented herein were obtained at the W. M. Keck Observatory from telescope time allocated to the National Aeronautics and Space Administration through the agency's scientific partnership with the California Institute of Technology and the University of California. The Observatory was made possible by the generous financial support of the W. M. Keck Foundation. The authors wish to recognize and acknowledge the very significant cultural role and reverence that the summit of Maunakea has always had within the Native Hawaiian community. We are most fortunate to have the opportunity to conduct observations from this mountain.

Time domain research by the University of Arizona team and D.J.S. is supported by National Science Foundation (NSF) grants 2308181, 2407566, and 2432036. 
JEA is supported by the international Gemini Observatory, a program of NSF NOIRLab, which is managed by the Association of Universities for Research in Astronomy (AURA) under a cooperative agreement with the U.S. National Science Foundation, on behalf of the Gemini partnership of Argentina, Brazil, Canada, Chile, the Republic of Korea, and the United States of America. KAB is supported by an LSST-DA Catalyst Fellowship; this publication was thus made possible through the support of Grant 62192 from the John Templeton Foundation to LSST-DA. E.R.B is supported by a Royal Society Dorothy Hodgkin Fellowship (grant no. DHF-R1-241114). N.F. acknowledges support from the National Science Foundation Graduate Research Fellowship Program under Grant No. DGE-2137419. Time-domain research by the University of California, Davis team and S.V. is supported by NSF grant AST-2407565. Supernova research at Rutgers University is support in part by NSF award AST-2407567. MS acknowledges funding from the Australian Research Council (ARC) Centre of Excellence CE230100016.
\end{acknowledgments}
 
\facilities{JWST (MIRI), Keck (LRIS), Gemini (GMOS), MAST (HLSP), MMT (Binospec, MMIRS)}
\setlength{\floatsep}{0pt} 

\software{\texttt{astropy} \citep{AstropyCollaboration2013, AstropyCollaboration2018, AstropyCollaboration2022}, emcee \citep{emcee}, LPipe \citep{Perley2019}, Matplotlib \citep{mpl}, Numpy \citep{numpy}, Photutils \citep{photutils}, \texttt{space\textunderscore phot} \citep{spacephot}, SEP \citep{Sextractor, SEP}, WebbPSF \citep{WebbPSF1, WebbPSF2}, WISeREP \citep{wiserep}, POTPyRI}
\vspace{-2em}
\startlongtable
\begin{deluxetable*}{lccccc}
\tabletypesize{\small}
\tablecaption{ JWST/MIRI photometry for the CCSN sample, ordered by phase relative to explosion. Reported upper limits correspond to $3\sigma$ limits. \label{tab:miri_photometry} }
\setlength{\tabcolsep}{10pt}
\tablehead{
\colhead{SN} & \colhead{Filter} & \colhead{MJD} & \colhead{Phase} & \colhead{$F_\nu$} & \colhead{AB Mag} \\ \colhead{} & \colhead{} & \colhead{(days)} & \colhead{(days)} & \colhead{($\mu$Jy)} & \colhead{(mag)}}
\startdata
SN2022wsp (+401 d) & F560W & 60255.80 & 401.0 & $34.37 \pm 0.31$ & $20.06 \pm 0.01$ \\
 & F770W & 60255.80 & 401.0 & $102.85 \pm 0.57$ & $18.87 \pm 0.01$ \\
 & F1000W & 60255.81 & 401.0 & $117.20 \pm 0.72$ & $18.73 \pm 0.01$ \\
 & F1130W & 60255.82 & 401.0 & $207.31 \pm 2.10$ & $18.11 \pm 0.01$ \\
 & F1280W & 60255.82 & 401.0 & $204.21 \pm 1.66$ & $18.12 \pm 0.01$ \\
 & F1500W & 60255.83 & 401.0 & $185.52 \pm 2.54$ & $18.23 \pm 0.02$ \\
 & F1800W & 60255.84 & 401.0 & $370.46 \pm 6.79$ & $17.48 \pm 0.02$ \\
 & F2100W & 60255.84 & 401.0 & $482.85 \pm 13.67$ & $17.19 \pm 0.03$ \\
 & F2550W & 60255.85 & 401.0 & $<488.40$ & $>17.18$ \\
\hline
SN2022acko (+405 d) & F560W & 60323.22 & 405.0 & $9.60 \pm 0.14$ & $21.44 \pm 0.02$ \\
 & F770W & 60323.23 & 405.0 & $18.83 \pm 0.16$ & $20.71 \pm 0.01$ \\
 & F1000W & 60323.24 & 405.0 & $9.65 \pm 0.31$ & $21.44 \pm 0.04$ \\
 & F1130W & 60323.24 & 405.0 & $11.60 \pm 0.84$ & $21.24 \pm 0.08$ \\
 & F1280W & 60323.24 & 405.0 & $9.53 \pm 0.64$ & $21.45 \pm 0.07$ \\
 & F1500W & 60323.25 & 405.0 & $5.52 \pm 0.83$ & $22.05 \pm 0.16$ \\
 & F1800W & 60323.26 & 405.0 & $14.31 \pm 1.81$ & $21.01 \pm 0.14$ \\
 & F2100W & 60323.26 & 405.0 & $<75.28$ & $>19.21$ \\
 & F2550W & 60323.27 & 405.0 & $<168.30$ & $>18.33$ \\
\hline
SN2024ggi (+660 d) & F560W & 61072.26 & 660.0 & $113 \pm 1.34$ & $18.76 \pm 0.01$ \\
 & F770W & 61072.27 & 660.0 & $263 \pm 1.70$ & $17.85 \pm 0.01$ \\
 & F1000W & 61072.27 & 660.0 & $449 \pm 3.34$ & $17.27 \pm 0.01$ \\
 & F1130W & 61072.28 & 660.0 & $522 \pm 7.70$ & $17.11 \pm 0.02$ \\
 & F1280W & 61072.29 & 660.0 & $560 \pm 6.06$ & $17.03 \pm 0.01$ \\
 & F1500W & 61072.29 & 660.0 & $611 \pm 8.06$ & $16.93 \pm 0.01$ \\
 & F1800W & 61072.30 & 660.0 & $724 \pm 17.22$ & $16.75 \pm 0.03$ \\
 & F2100W & 61072.30 & 660.0 & $680 \pm 30.38$ & $16.82 \pm 0.05$ \\
 & F2550W & 61072.31 & 660.0 & $668 \pm 126.00$ & $16.84 \pm 0.21$ \\
\hline
SN2022jox (+701 d) & F560W & 60409.12 & 701.0 & $3.49 \pm 0.14$ & $22.54 \pm 0.04$ \\
 & F770W & 60409.12 & 701.0 & $7.28 \pm 0.16$ & $21.74 \pm 0.02$ \\
 & F1000W & 60409.13 & 701.0 & $27.38 \pm 0.39$ & $20.31 \pm 0.02$ \\
 & F1130W & 60409.14 & 701.0 & $34.34 \pm 1.22$ & $20.06 \pm 0.04$ \\
 & F1280W & 60409.14 & 701.0 & $31.65 \pm 1.07$ & $20.15 \pm 0.04$ \\
 & F1500W & 60409.15 & 701.0 & $42.11 \pm 2.03$ & $19.84 \pm 0.05$ \\
 & F1800W & 60409.16 & 701.0 & $54.38 \pm 6.04$ & $19.56 \pm 0.12$ \\
 & F2100W & 60409.16 & 701.0 & $42.94 \pm 13.02$ & $19.82 \pm 0.33$ \\
 & F2550W & 60409.17 & 701.0 & $82.58 \pm 15.69$ & $19.11 \pm 0.21$ \\
\hline
SN2021yja (+859 d) & F560W & 60323.29 & 859.0 & $7.44 \pm 0.14$ & $21.72 \pm 0.02$ \\
 & F770W & 60323.29 & 859.0 & $16.11 \pm 0.18$ & $20.88 \pm 0.01$ \\
 & F1000W & 60323.30 & 859.0 & $46.05 \pm 0.48$ & $19.74 \pm 0.01$ \\
 & F1130W & 60323.31 & 859.0 & $53.45 \pm 1.41$ & $19.58 \pm 0.03$ \\
 & F1280W & 60323.31 & 859.0 & $48.63 \pm 1.17$ & $19.68 \pm 0.03$ \\
 & F1500W & 60323.32 & 859.0 & $58.92 \pm 2.19$ & $19.47 \pm 0.04$ \\
 & F1800W & 60323.32 & 859.0 & $95.40 \pm 6.51$ & $18.95 \pm 0.07$ \\
 & F2100W & 60323.33 & 859.0 & $100.21 \pm 14.29$ & $18.90 \pm 0.16$ \\
 & F2550W & 60323.34 & 859.0 & $101.21 \pm 25.84$ & $18.89 \pm 0.28$ \\
\hline
SN2023ixf (+992 d) & F560W & 61075.54 & 992.0 & $595 \pm 4.03$ & $16.96 \pm 0.01$ \\
 & F770W & 61075.54 & 992.0 & $1143 \pm 4.08$ & $16.25 \pm 0.01$ \\
 & F1000W & 61075.55 & 992.0 & $3148 \pm 21.88$ & $15.16 \pm 0.01$ \\
 & F1130W & 61075.55 & 992.0 & $3881 \pm 30.66$ & $14.93 \pm 0.01$ \\
 & F1280W & 61075.56 & 992.0 & $3343 \pm 27.94$ & $15.09 \pm 0.01$ \\
 & F1500W & 61075.56 & 992.0 & $3289 \pm 28.38$ & $15.11 \pm 0.01$ \\
 & F1800W & 61075.57 & 992.0 & $4065 \pm 66.92$ & $14.88 \pm 0.0$ \\
 & F2100W & 61075.58 & 992.0 & $4040 \pm 65.72$ & $14.88 \pm 0.02$ \\
 & F2550W & 61075.59 & 992.0 & $3627 \pm 127.60$ & $15.00 \pm 0.04$ \\
\hline
SN2021gmj (+1120 d) & F560W & 60412.20 & 1120.0 & $3.31 \pm 0.15$ & $22.60 \pm 0.05$ \\
 & F770W & 60412.21 & 1120.0 & $18.99 \pm 0.27$ & $20.70 \pm 0.02$ \\
 & F1000W & 60412.22 & 1120.0 & $13.76 \pm 0.49$ & $21.05 \pm 0.04$ \\
 & F1130W & 60412.23 & 1120.0 & $47.39 \pm 1.71$ & $19.71 \pm 0.04$ \\
 & F1280W & 60412.23 & 1120.0 & $40.90 \pm 1.25$ & $19.87 \pm 0.03$ \\
 & F1500W & 60412.24 & 1120.0 & $52.37 \pm 1.12$ & $19.60 \pm 0.02$ \\
 & F1800W & 60412.25 & 1120.0 & $83.61 \pm 6.61$ & $19.09 \pm 0.09$ \\
 & F2100W & 60412.25 & 1120.0 & $126.84 \pm 6.15$ & $18.64 \pm 0.05$ \\
 & F2550W & 60412.26 & 1120.0 & $<110.54$ & $>18.79$ \\
\hline
SN2020jfo (+1471 d) & F560W & 60445.56 & 1471.0 & $1.22 \pm 0.16$ & $23.68 \pm 0.14$ \\
 & F770W & 60445.57 & 1471.0 & $11.00 \pm 0.21$ & $21.30 \pm 0.02$ \\
 & F1000W & 60445.57 & 1471.0 & $7.59 \pm 0.40$ & $21.70 \pm 0.06$ \\
 & F1130W & 60445.57 & 1471.0 & $28.01 \pm 1.45$ & $20.28 \pm 0.06$ \\
 & F1280W & 60445.58 & 1471.0 & $17.81 \pm 1.05$ & $20.77 \pm 0.06$ \\
 & F1500W & 60445.59 & 1471.0 & $16.99 \pm 1.85$ & $20.82 \pm 0.12$ \\
 & F1800W & 60445.60 & 1471.0 & $25.93 \pm 3.73$ & $20.37 \pm 0.16$ \\
 & F2100W & 60445.60 & 1471.0 & $40.96 \pm 6.91$ & $19.87 \pm 0.18$ \\
 & F2550W & 60445.61 & 1471.0 & $<224.46$ & $>18.02$ \\
\hline
SN2018cuf (+1929 d) & F560W & 60221.75 & 1929.0 & $0.55 \pm 0.14$ & $24.56 \pm 0.28$ \\
 & F770W & 60221.76 & 1929.0 & $4.20 \pm 0.16$ & $22.34 \pm 0.04$ \\
 & F1000W & 60221.77 & 1929.0 & $2.06 \pm 0.46$ & $23.11 \pm 0.24$ \\
 & F1130W & 60221.78 & 1929.0 & $5.37 \pm 1.34$ & $22.08 \pm 0.27$ \\
 & F1280W & 60221.78 & 1929.0 & $5.19 \pm 1.00$ & $22.11 \pm 0.21$ \\
 & F1500W & 60221.78 & 1929.0 & $4.60 \pm 0.37$ & $22.24 \pm 0.09$ \\
 & F1800W & 60221.79 & 1929.0 & $10.13 \pm 2.66$ & $21.39 \pm 0.29$ \\
 & F2100W & 60221.80 & 1929.0 & $33.46 \pm 9.69$ & $20.09 \pm 0.31$ \\
 & F2550W & 60221.80 & 1929.0 & $<552.79$ & $>17.04$ \\
\hline
SN2017eaw (+2330 d) & F560W & 60213.39 & 2330.0 & $2.85 \pm 0.13$ & $22.76 \pm 0.05$ \\
 & F770W & 60213.39 & 2330.0 & $6.84 \pm 0.17$ & $21.81 \pm 0.03$ \\
 & F1000W & 60213.40 & 2330.0 & $36.89 \pm 0.45$ & $19.98 \pm 0.01$ \\
 & F1130W & 60213.41 & 2330.0 & $36.43 \pm 1.29$ & $20.00 \pm 0.04$ \\
 & F1280W & 60213.41 & 2330.0 & $31.93 \pm 1.05$ & $20.14 \pm 0.04$ \\
 & F1500W & 60213.42 & 2330.0 & $44.33 \pm 1.90$ & $19.78 \pm 0.05$ \\
 & F1800W & 60213.43 & 2330.0 & $99.67 \pm 5.84$ & $18.90 \pm 0.06$ \\
 & F2100W & 60213.43 & 2330.0 & $102.81 \pm 14.25$ & $18.87 \pm 0.15$ \\
 & F2550W & 60213.44 & 2330.0 & $78.51 \pm 19.08$ & $19.16 \pm 0.26$ \\
\hline
SN2017gmr (+2335 d) & F560W & 60334.33 & 2335.0 & $2.37 \pm 0.12$ & $22.96 \pm 0.06$ \\
 & F770W & 60334.34 & 2335.0 & $5.70 \pm 0.16$ & $22.01 \pm 0.03$ \\
 & F1000W & 60334.35 & 2335.0 & $29.81 \pm 0.38$ & $20.21 \pm 0.01$ \\
 & F1130W & 60334.35 & 2335.0 & $24.08 \pm 0.95$ & $20.45 \pm 0.04$ \\
 & F1280W & 60334.36 & 2335.0 & $19.09 \pm 0.74$ & $20.70 \pm 0.04$ \\
 & F1500W & 60334.37 & 2335.0 & $22.29 \pm 1.18$ & $20.53 \pm 0.06$ \\
 & F1800W & 60334.37 & 2335.0 & $37.03 \pm 2.83$ & $19.98 \pm 0.08$ \\
 & F2100W & 60334.38 & 2335.0 & $47.15 \pm 3.72$ & $19.72 \pm 0.09$ \\
 & F2550W & 60334.38 & 2335.0 & $<207.09$ & $>18.11$
\enddata
\end{deluxetable*}
\startlongtable
\begin{deluxetable*}{llcc}
\tabletypesize{\small}
\tablecaption{ Best fit dust parameters from the MCMC modeling of the JWST/MIRI SEDs. The reported values correspond to the median of the posterior distribution, with uncertainties given by the 16th and 84th percentiles. } \label{tab:fit_params}
\setlength{\tabcolsep}{10pt}
\tablehead{
  \colhead{SN} & \colhead{Parameter} & \colhead{Prior} & \colhead{All Filters} \\
  \colhead{}  & \colhead{} & \colhead{} & \colhead{ (exc.\ F770W, F1130W, F1280W)}
}
\startdata
%
\multicolumn{4}{l}{\textbf{Si\_BB} (Si; $\rho_{\rm Si}=3.3$~g~cm$^{-3}$)} \\
\hline
SN\,2022wsp ($+401$\,d)  & $T_{\rm BB}$\,(kK)                & Uniform$(1.0,\,2.0)$               & $1.51^{+0.53}_{-0.32}$  \\
             & $R_{\rm BB}$\,($10^3\,R_\odot$)   & Log-Uniform$(1,\,10^3)$            & $12.9^{+4.0}_{-3.2}$  \\
             & $T_{\rm dust}$\,(kK)               & Uniform$(0.05,\,0.3)$              & $0.166^{+0.003}_{-0.003}$ \\
             & $M_{\rm dust}$\,($M_\odot$)        & Log-Uniform$(10^{-8},\,1)$        & $1.32^{+0.10}_{-0.08}\times10^{-2}$ \\
\hline
SN\,2022acko ($+405$\,d)  & $T_{\rm BB}$\,(kK)                & Uniform$(1.0,\,2.0)$               & $1.36^{+0.74}_{-0.29}$  \\
             & $R_{\rm BB}$\,($10^3\,R_\odot$)   & Log-Uniform$(1,\,10^3)$            & $7.30^{+2.4}_{-2.5}$  \\
             & $T_{\rm dust}$\,(kK)               & Uniform$(0.05,\,0.3)$              & $0.183^{+0.008}_{-0.009}$ \\
             & $M_{\rm dust}$\,($M_\odot$)        & Log-Uniform$(10^{-8},\,1)$        & $1.95^{+0.62}_{-0.52}\times10^{-4}$ \\
\hline
SN\,2024ggi ($+660$\,d)  & $T_{\rm BB}$\,(kK)                & Uniform$(0.01,\,2.0)$               & $0.404^{+0.004}_{-0.004}$  \\
             & $R_{\rm BB}$\,($10^3\,R_\odot$)   & Log-Uniform$(1,\,10^3)$            & $99.35^{+0.49}_{-1.05}$  \\
             & $T_{\rm dust}$\,(kK)               & Uniform$(0.05,\,0.3)$              & $0.182^{+0.004}_{-0.002}$ \\
             & $M_{\rm dust}$\,($M_\odot$)        & Log-Uniform$(10^{-8},\,1)$        & $1.44^{+0.07}_{-0.07}\times10^{-3}$ \\
\hline
SN\,2022jox ($+701$\,d)  & $T_{\rm BB}$\,(kK)                & Uniform$(1.0,\,2.0)$               & $1.73^{+0.50}_{-0.48}$  \\
             & $R_{\rm BB}$\,($10^3\,R_\odot$)   & Log-Uniform$(1,\,10^3)$            & $7.0^{+2.9}_{-1.4}$  \\
             & $T_{\rm dust}$\,(kK)               & Uniform$(0.05,\,0.3)$              & $0.198^{+0.003}_{-0.005}$ \\
             & $M_{\rm dust}$\,($M_\odot$)        & Log-Uniform$(10^{-8},\,1)$        & $4.86^{+0.58}_{-0.46}\times10^{-3}$ \\
\hline
SN\,2021yja ($+859$\,d)  & $T_{\rm BB}$\,(kK)                & Uniform$(1.0,\,2.0)$               & $1.70^{+0.55}_{-0.51}$  \\
             & $R_{\rm BB}$\,($10^3\,R_\odot$)   & Log-Uniform$(1,\,10^3)$            & $6.4^{+3.0}_{-1.4}$  \\
             & $T_{\rm dust}$\,(kK)               & Uniform$(0.05,\,0.3)$              & $0.192^{+0.003}_{-0.005}$ \\
             & $M_{\rm dust}$\,($M_\odot$)        & Log-Uniform$(10^{-8},\,1)$        & $2.81^{+0.32}_{-0.24}\times10^{-3}$ \\
\hline
%
\multicolumn{4}{l}{\textbf{CSi\_thin} (Si + amC; $\rho_{\rm Si}=3.3$, $\rho_{\rm amC}=2.5$~g~cm$^{-3}$)} \\
\hline
SN\,2021gmj ($+1120$\,d)  & $T_{\rm hot}$\,(kK)               & Uniform$(1.0,\,1.5)$               & $1.71^{+0.20}_{-0.24}$  \\
             & $M_{\rm hot}$\,($M_\odot$)        & Log-Uniform$(10^{-8},\,10^{-6})$  & $6.21^{+2.22}_{-1.20}\times10^{-8}$  \\
             & $T_{\rm cool}$\,(kK)              & Uniform$(0.05,\,0.2)$              & $0.130^{+0.005}_{-0.005}$ \\
             & $M_{\rm cool}$\,($M_\odot$)       & Log-Uniform$(10^{-5},\,1)$        & $1.231^{+0.092}_{-0.089}\times10^{-2}$ \\
\hline
SN\,2020jfo ($+1471$\,d)  & $T_{\rm hot}$\,(kK)               & Uniform$(1.0,\,1.5)$               & $1.67^{+0.22}_{-0.24}$  \\
             & $M_{\rm hot}$\,($M_\odot$)        & Log-Uniform$(10^{-8},\,10^{-6})$  & $1.45^{+0.55}_{-0.33}\times10^{-8}$  \\
             & $T_{\rm cool}$\,(kK)              & Uniform$(0.05,\,0.2)$              & $0.152^{+0.003}_{-0.003}$ \\
             & $M_{\rm cool}$\,($M_\odot$)       & Log-Uniform$(10^{-5},\,1)$        & $9.03^{+1.65}_{-1.48}\times10^{-4}$ \\
\hline
%
\multicolumn{4}{l}{\textbf{C2Si\_thin} (2Gr + Si; $\rho_{\rm Gr}=2.26$, $\rho_{\rm Si}=3.3$~g~cm$^{-3}$)} \\
\hline
SN\,2023ixf ($+992$\,d)  & $T_{\rm C,hot}$\,(kK)              & Uniform$(0.7,\,1.0)$               & $0.985^{+0.01}_{-0.02}$  \\
             & $M_{\rm C,hot}$\,($M_\odot$)      & Log-Uniform$(10^{-15},\,1)$       & $1.15^{+0.23}_{-0.16}\times10^{-6}$  \\
             & $T_{\rm C,cool}$\,(kK)            & Uniform$(0.1,\,0.4)$               & $0.278^{+0.002}_{-0.002}$  \\
             & $M_{\rm C,cool}$\,($M_\odot$)     & Log-Uniform$(10^{-15},\,1)$       & $6.05^{+0.32}_{-0.32}\times10^{-3}$  \\
             & $T_{\rm Si}$\,(kK)                & Uniform$(0.01,\,0.3)$              & $0.223^{+0.006}_{-0.005}$  \\
             & $M_{\rm Si}$\,($M_\odot$)         & Log-Uniform$(10^{-15},\,1)$       & $1.30^{+0.26}_{-0.24}\times10^{-3}$  \\
\hline
%
\multicolumn{4}{l}{\textbf{CSi\_thin} (Si + Gr; $\rho_{\rm Si}=3.3$, $\rho_{\rm Gr}=2.26$~g~cm$^{-3}$)} \\
\hline
SN\,2018cuf ($+1929$\,d)  & $T_{\rm hot}$\,(kK)               & Uniform$(1.0,\,1.5)$               & $1.57^{+0.30}_{-0.36}$  \\
             & $M_{\rm hot}$\,($M_\odot$)        & Log-Uniform$(10^{-8},\,10^{-6})$  & $8.84^{+6.78}_{-3.25}\times10^{-8}$  \\
             & $T_{\rm cool}$\,(kK)              & Uniform$(0.05,\,0.2)$              & $0.141^{+0.005}_{-0.005}$ \\
             & $M_{\rm cool}$\,($M_\odot$)       & Log-Uniform$(10^{-5},\,1)$        & $4.18^{+1.27}_{-1.01}\times10^{-3}$ \\
\hline
%
\multicolumn{4}{l}{\textbf{2Si\_thin} (2Si; $\rho_{\rm Si}=3.3$~g~cm$^{-3}$)} \\
\hline
SN\,2017eaw ($+2330$\,d)  & $T_{\rm hot}$\,(kK)               & Uniform$(1.0,\,1.5)$               & $1.73^{+0.03}_{-0.03}$  \\
             & $M_{\rm hot}$\,($M_\odot$)        & Log-Uniform$(10^{-8},\,10^{-6})$  & $5.30^{+0.29}_{-0.27}\times10^{-8}$  \\
             & $T_{\rm cool}$\,(kK)              & Uniform$(0.05,\,0.2)$              & $0.149^{+0.002}_{-0.002}$ \\
             & $M_{\rm cool}$\,($M_\odot$)       & Log-Uniform$(10^{-5},\,1)$        & $7.28^{+0.71}_{-0.67}\times10^{-4}$ \\
\hline
SN\,2017gmr ($+2335$\,d)  & $T_{\rm hot}$\,(kK)               & Uniform$(1.0,\,1.5)$               & $1.58^{+0.28}_{-0.31}$  \\
             & $M_{\rm hot}$\,($M_\odot$)        & Log-Uniform$(10^{-8},\,10^{-6})$  & $3.26^{+2.02}_{-0.89}\times10^{-7}$  \\
             & $T_{\rm cool}$\,(kK)              & Uniform$(0.05,\,0.2)$              & $0.174^{+0.004}_{-0.006}$ \\
             & $M_{\rm cool}$\,($M_\odot$)       & Log-Uniform$(10^{-5},\,1)$        & $9.89^{+1.71}_{-1.31}\times10^{-4}$ \\
\hline
\enddata
\end{deluxetable*}

\vspace{-5em}
\appendix
\vspace{-2em}
\section{Host environment complexity and PAH and [Ne II] sensitive bands}
\label{appendix:host_complexity}

Broad band MIR photometry in crowded, star-forming host environments is vulnerable to contamination from diffuse emission, line emission, and spatially structured background residuals. We therefore exclude F770W, F1130W, and F1280W from the dust-model fitting for all objects in the sample, regardless of whether any individual source appears to show an obvious excess in those bands. This choice enforces a uniform wavelength baseline across the sample.

\begin{figure}[htp]
    \centering
    \includegraphics[width=\textwidth]{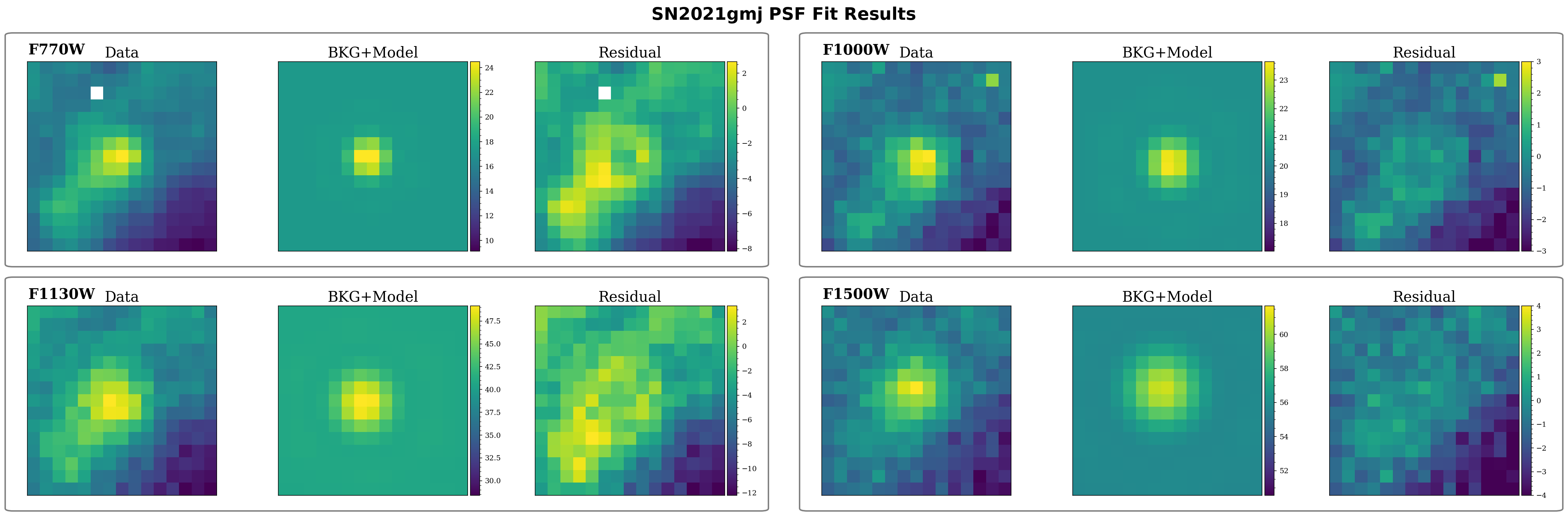}
    \caption{PSF-fit comparison for SN~2021gmj in the F770W and F1130W bands (left) and the F1000W and F1500W bands (right). Structured residuals in F770W (the white square marks a bad pixel) and F1130W reveal spatially varying host emission not captured by the model, while the residuals in F1000W and F1500W are relatively uniform, although trace residuals remain. }
    \label{fig:sn2021gmj_residuals}
\end{figure}

The motivation for treating these filters separately is physical rather than purely empirical. F770W and F1130W are PAH sensitive because they encompass the prominent 7.7 and 11.3 $\mu$m PAH complexes, respectively, while F1280W is [\ion{Ne}{2}] sensitive and can therefore be affected by narrow spectral structure that broad band imaging cannot disentangle from the underlying continuum. SN~2021gmj illustrates this issue particularly well as seen in Figure \ref{fig:sn2021gmj_residuals}, and is used as an illustrative example. The PSF fit residuals in F770W and F1130W show pronounced structure across the local environment, demonstrating that the source is embedded in a non uniform background that is not fully captured by the adopted point source function plus background model, as compared to the other bands (F1000W, F1500W). The residual pattern is not consistent with a clean isolated point source on a smooth field, especially in the PAH-sensitive bands, and therefore the extracted fluxes in these filters cannot be interpreted as unambiguous measurements of the SN continuum. 

Accordingly, the apparent excesses in these bands are best regarded as evidence of local host environment complexity, not as proof of SN associated PAH or line emission, although we cannot rule out supernova-related emission. Mid-infrared spectroscopy would be required to separate continuum emission from PAH features and any narrow line contribution \citep{Shrestha2026}. In the absence of such spectral information, we do not include these bands into our continuum dust fits.  We discuss the implications for our fitting in Appendix \ref{appendix:filter_robustness}.

\section{Implications of the dust fits to filter exclusion}
\label{appendix:filter_robustness}

\begin{figure}[ht!]
    \centering
    \includegraphics[width=0.9\textwidth]{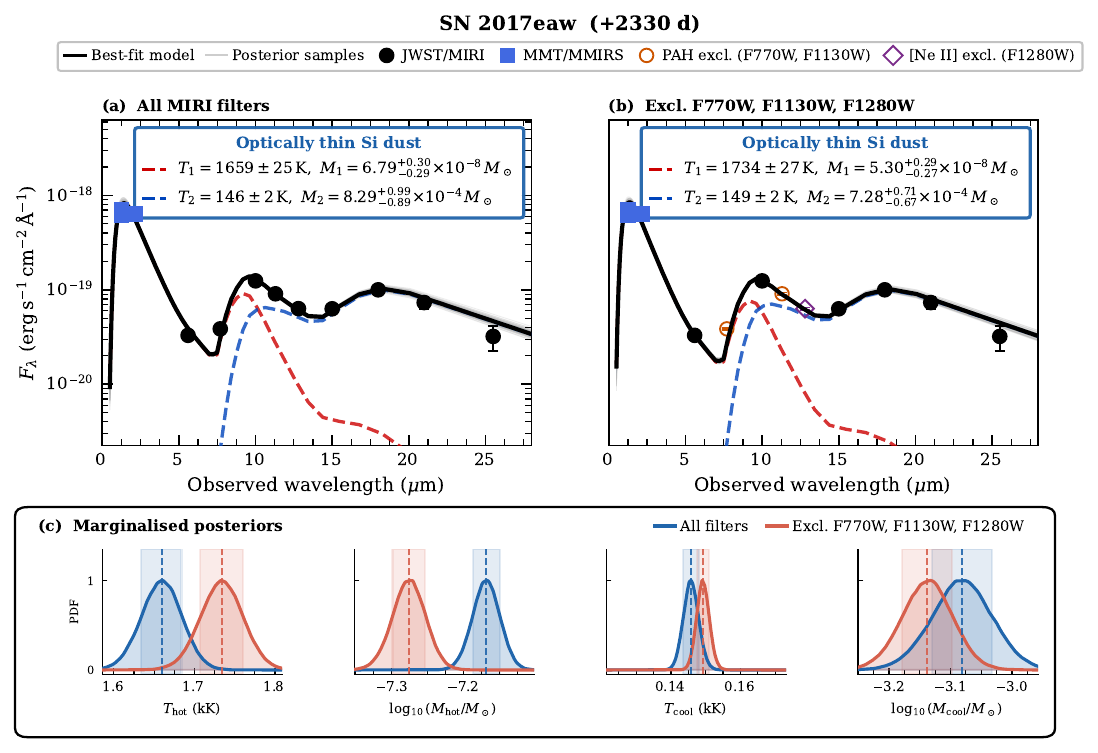}
    \caption{Comparison of dust model fits to SN~2017eaw using a) all MIRI filters and b) after excluding F770W, F1130W, and F1280W. The best-fit model and marginalized posterior distributions are nearly unchanged (panel c), indicating that the dust temperature and mass estimates are robust even after removal of these bands.}
    \label{fig:sn2017eaw_filter_comparison}
\end{figure}

To verify that removing the PAH-sensitive and \ion{Ne}{2}-sensitive bands does not materially alter the inferred continuum dust properties, we compare SN~2017eaw fits obtained with all MIRI filters to fits obtained after excluding F770W, F1130W, and F1280W. This object serves as the validation case for the sample as a whole: it demonstrates whether excluding F770W, F1130W, and F1280W changes the derived continuum parameters in any meaningful way. SN~2017eaw was chosen because its MIRI SED shows no apparent excess in F770W, F1130W, or F1280W, making it a suitable baseline for this test. The conclusion is that the removal of these filters does not significantly alter the fit results, as shown in Figure \ref{fig:sn2017eaw_filter_comparison}. The best-fit values change only slightly between the two fits, and the marginalized posterior distributions for both the hot and cool dust components remain nearly coincident and within each other's uncertainities. Using all filters, the best-fit hot component is $T_1 = 1659 \pm 25$ K with
$M_1 = 6.79^{+0.30}_{-0.29} \times 10^{-8}\,M_\odot$, and the cool component is $T_2 = 146 \pm 2$ K with $M_2 = 8.29^{+0.99}_{-0.89} \times 10^{-4}\,M_\odot$. After removing F770W, F1130W, and F1280W, the corresponding values are $T_1 = 1734 \pm 26$ K, $M_1 = 5.30^{+0.29}_{-0.27} \times 10^{-8}\,M_\odot$, $T_2 = 149 \pm 2$ K, and $M_2 = 7.28^{+0.71}_{-0.67} \times 10^{-4}\,M_\odot$. These differences are within the 68\% credible intervals and do not alter the physical interpretation of the fit. 

\vspace{-3em}
\bibliography{references}{}
\bibliographystyle{aasjournalv7}

\end{document}